\documentclass[11pt,letterpaper]{article}
\usepackage{putex}
\usepackage{graphicx}
\usepackage{latexsym,amsmath,amsfonts,amssymb,amsthm}
\usepackage{bbm}
\usepackage{cite}
\usepackage{indentfirst}
\usepackage{booktabs,array}
\usepackage{placeins}
\usepackage{mathtools}
\usepackage[dvipsnames]{xcolor}
\usepackage{tikz}
\usepackage{enumitem}
\usepackage{setspace}
\usetikzlibrary{decorations.markings}
\usepackage{graphicx}
\usepackage[margin=10pt,font=small,labelfont=bf]{caption}
\usepackage{subcaption}
\usepackage{xcolor}
\usepackage{float}
\usepackage{microtype}
\usepackage{hyperref}

\setlist{itemsep=2pt plus 1pt minus 1pt, topsep=2pt plus 1pt minus 1pt}

\renewenvironment{thebibliography}[1]{%
\begin{oldthebibliography}{#1}%
\setlength\itemsep{-2mm}%
}%
{%
\end{oldthebibliography}%
}

\numberwithin{equation}{section}

\begin{document}


\title{\begin{LARGE}
Intersecting Surface defects and 3d Superconformal indices\newline
\end{LARGE}}

\authors{Junfeng Liu, Yiwen Pan, Hong-Hao Zhang
\medskip\medskip\medskip\medskip
 }

\institution{SYSU}{${}$
School of Physics, Sun Yat-Sen University, Guangzhou 510275, China}

\abstract{\begin{onehalfspace}{
  We compute the 3d $\mathcal{N} = 2$ superconformal indices for 3d/1d coupled systems, which arise as the worldvolume theories of intersecting surface defects engineered by Higgsing 5d $\mathcal{N} = 1$ gauge theories. We generalize some known 3d dualities, including non-Abelian 3d mirror symmetry and 3d/3d correspondence, to some of the simple 3d/1d coupled systems. Finally we propose a $q$-Virasoro construction for the superconformal indices.
}\end{onehalfspace}}

\preprint{}
\setcounter{page}{0}
\maketitle


{
\setcounter{tocdepth}{2}
\setlength\parskip{-0.7mm}
\tableofcontents
}

\section{Introduction}

Since the seminal work by Pestun \cite{Pestun:2007rz}, numerous exact results have been derived using the technique of supersymmetric localization for supersymmetric theories in different dimensions. Two simplest quantities that admit localization computations are superconformal indices and sphere partition functions \footnote{See also some localization computations performed on manifolds with boundaries \cite{Kimura:2018axa,Longhi:2019hdh,Sugiyama:2020uqh,Yoshida:2014ssa}, which are closely related to the factorization \cite{Nieri:2015yia,Pasquetti:2011fj,Hwang:2015wna} of partition functions and indices.} \cite{Benini:2012ui,Doroud:2012xw,Hama:2010av,Hama:2011ea,Fujitsuka:2013fga,Benvenuti:2011ga,Benini:2013yva,Hama:2012bg,Kapustin:2011jm,Pan:2015hza,Chen:2015fta,Kallen:2012va,Kallen:2012cs,Minahan:2015jta}, which can be further decorated with local BPS operators \cite{Dedushenko:2016jxl,Dedushenko:2017avn,Pan:2017zie,Pan:2019bor,Dedushenko:2019yiw,Pan:2019shz,Dedushenko:2019mnd,Panerai:2020boq,Oh:2019bgz,Jeong:2019pzg} or non-local BPS defects \cite{Gorsky:2017hro,Drukker:2012sr,Giombi:2009ek,Giombi:2009ds,Assel:2015oxa,Lamy-Poirier:2014sea}. Among these BPS insertions are the particularly interesting codimension-two defects, which are usually referred to simply as surface defects. They are considered important tools to identify phases of quantum field theories \cite{Gukov:2013zka}.

The field-theoretic construction of surface defects usually falls into three, probably overlapping if properly identified, categories. One is by prescribing some symmetry-preserving singular behavior of the fundamental fields in the theory near the locus where the defect resides \cite{Gukov:2006jk}. A second approach is to place a supersymmetric theory on the locus that will couple to the bulk theory in such a way that a certain amount of supersymmetry is preserved. A third way is to trigger an RG-flow by giving a position dependent vacuum expectation value to some operator and then look at the resulting theory in the IR \cite{Gaiotto:2012xa,Gaiotto:2014ina}. In simplest cases, the locus of the defect is a submanifold in the bulk space, while generally it can be the union of multiple intersecting submanifolds. In the latter cases, the second approach above would require one to place two supersymmetric theories on the two codimension-two submanifolds, and some further supersymmetric theories on the intersection of higher codimensions, while all these lower-dimensional theories are further coupled to the original bulk theory in a supersymmetric manner, forming a $n$d/$(n-2)$d/$(n-4)$d coupled system. We shall refer to these general defects as intersecting surface defects \cite{Pan:2016fbl,Gomis:2016ljm,Nieri:2018ghd,Nieri:2018pev,Jeong:2020uxz,Ferrari:2020avq}.

The sphere partition functions in the presence of a wide variety of surface defects, including intersecting surface defects, are computed and shown to participate in different dualities. For example, surface defects in a 4d $\mathcal{N} = 2$ SCFTs engineered by coupling a class of 2d gauged linear sigma models to the bulk are shown to be dual to general degenerate vertex operators in the Liouville/Toda theory through the AGT-duality \cite{Alday:2009aq,Gomis:2014eya,Bonelli:2011wx,Alday:2009fs,Gomis:2016ljm}. A similar 5d uplift to $S^5$-partition function and to 5d index were also discussed \cite{Nieri:2013vba}. When the bulk theory is a free theory, the $(n - 2)$d part (or the $(n-2)$d/$(n-4)$d part in the intersecting case) of the full theory can be isolated and one can study dualities they enjoy. For instance, as defect worldvolume theories in a 5d $\mathcal{N} = 1$ theory, a class of (intersecting) 3d $\mathcal{N} = 2$ SQCDA with one flavor are shown to enjoy 3d $\mathcal{N} = 2$ mirror symmetry that descends from the fiber-base duality \cite{Aprile:2018oau,Nieri:2018pev,Benvenuti:2016wet}. The list goes on.

In this paper, we continue to investigate the 3d $\mathcal{N} = 2$ superconformal indices of 3d/1d coupled systems, viewed as surface defects in a 5d $\mathcal{N} = 1$ theory. Starting with the 5d index of the standard $U(N)$ SQCD, we perform the Higgsing procedure to extract the indices of the resulting 3d/1d coupled systems. Viewing the 5d index as a $S^4_{\epsilon_1, \epsilon_2} \times S^1$ partition function, the 3d/1d coupled systems are built out of two 3d $\mathcal{N} = 2$ $U(n^\text{L,R})$ gauge theories on $S^2_\text{L,R} \times S^1 \subset S^4\times S^1$ which further interact with some 1d bifundamental chiral multiplets on the intersection $S^1$. These 3d/1d systems if circle-reduced to 2d/0d would correspond to the general degenerate Liouville/Toda vertex operators labeled by a pair of symmetric representations.

3d $\mathcal{N} = 2$ gauge theories (possibly with Chern-Simons term) are known to enjoy 3d mirror symmetry \cite{Benvenuti:2016wet,Cheng:2020zbh}, which is a generalization to the $\mathcal{N} = 4$ version \cite{Intriligator:1996ex,Kapustin:1999ha}. For example, a standard mirror symmetry is between $U(1)_{\frac{1}{2}}$ theory with one fundamental chiral and the free theory with one chiral. When we place two such theories on $S^2_\text{L} \times S^1 \cup S^2_\text{R} \times S^1$ and couple them to a bifundamental chiral on the intersection, we show that the mirror symmetry generalizes. Similarly, if one starts with a 5d $U(1)$ SQED, it is well-known that it enjoys a fiber-base duality with some 5d free theory. The duality is expected to descend to 3d $\mathcal{N} = 2$ mirror symmetry between 3d/1d coupled systems, generalizing the usual duality between SQED and the XYZ model. This has been checked by comparing the $S^3_\text{L} \cup S^3_\text{R}$ partition functions, and in this paper we provide further evidence by also computing the superconformal indices for such 3d/1d coupled systems.

A class of 3d $\mathcal{N} = 2$ theories $\mathcal{T}[M]$ can be engineered by a twisted compactification from 6d on a three manifold $M$ \cite{Dimofte:2011py,Dimofte:2010tz,Dimofte:2011ju,Gukov:2016gkn}, similar to the class-$\mathcal{S}$ construction. In particular, the superconformal indices of $\mathcal{T}[M]$ are known to compute the Chern-Simons partition functions on $S^3$, which is one simple entry in the 3d/3d correspondence. It is therefore natural to ask if it generalizes to intersecting 3d $\mathcal{N} = 2$ theories. We report an equality at the level of $D^2 \times S^1$ partition function of a simplest theory of intersecting SQEDs and a matrix integral, which could be a generalization to the known duality. However, the precise physical interpretation remain unclear and is left for future study.

Partition functions and indices of 3d $\mathcal{N} = 2$ theories can be constructed using the screening charges $q$-Virasoro algebras \cite{Nedelin:2016gwu}, and therefore they sit in the kernel of some differential operators if proper formal variables are included into the partition functions and indices \cite{Nedelin:2015mio,Nedelin:2016gwu}. Generalization to higher dimension is also possible \cite{Lodin:2017lrc}. Another generalization, refereed to as a modular triple \cite{Nieri:2017vrb}, to accommodate partition functions of intersecting theories on $S^3_{(1)} \cup S^3_{(2)} (\cup S^3_{(3)})$ was proposed and was shown to be the the only solutions under the requirement that all the participating screening charges commute with all the participating $q$-Virasoro stress tensors. In this paper we propose a similar construction for intersecting superconformal indices, and also argue the uniqueness of the construction.

The organization of the paper is as follows. In section 2 we will review the Higgsing procedure that engineers a type of surface defects that we will be studying, as well as the corresponding brane construction. In section 3, we apply the procedure to 5d $\mathcal{N} =1$ SQCD and extract the 3d $\mathcal{N} = 2$ indices of the worldvolume theories as 3d/1d coupled systems. In section 4, we investigate some possible dualities that these types of theories enjoy, including 3d $\mathcal{N} = 2$ mirror symmetry and 3d/3d correspondence. In section 5, we propose a $q$-Virasoro construction of the superconformal indices for the 3d/1d coupled systems and argue its uniqueness.

\section{Higgsing and surface defects}

In this section we review an approach to constructing a class of codimension-two BPS defects in 5d $\mathcal{N} = 1$ (or 4d $\mathcal{N} = 2$) supersymmetric gauge theories, referred to as the Higgsin procedure following \cite{Gaiotto:2012xa}.

Consider a theory $\mathcal{T}$ with a flavor symmetry subgroup $SU(N)$. One can bring in an additional $N^2$ free hypermultiplets with flavor symmetry subgroup $SU(N) \times SU(N) \times U(1)$ and gauge the diagonal subgroup of the $SU(N)$ from $\mathcal{T}$ and one $SU(N)$ factor from the free hypermultiplets. The resulting theory will be called $\tilde{\mathcal{T}}$ and has an additional $U(1)$ flavor symmetry, compared with $\mathcal{T}$. One can then turn on the vacuum expectation value of a baryonic Hiiggs branch operator associated to the $U(1)$ factor and trigger an RG-flow. In particular, the expectation value can be position dependent with a core where the expectation vanishes. If the position dependence is trivial, in the IR one recovers the original theory $\mathcal{T}$, while for non-trivial dependence, one recovers the original theory coupled to a surface defect of codimension-two sitting at the core. Note that the core is not necessarily a submanifold, but in general could be the union of two intersecting submanifolds. In such case, we will refer to the surface defects as an intersecting surface defect.

The above construction can be further visualized in detail at the level of sphere partition functions or superconformal indices. For the purpose of this paper, consider the case of the superconformal index of a 5d $\mathcal{N} = 1$ gauge theory. One can compute the index of $\tilde{\mathcal{T}}$ which will be a function of the $U(1)$ flavor fugacity $b$. The index has poles at special values of $b$, which correspond to the values of $b$ such that the integration contour is pinched by poles of the integrand therein. More concretely, if one starts with $\mathcal{T}$ as $N^2$ free hypermultiplets, then $\tilde{\mathcal{T}}$ will be the standard $SU(N)$ SQCD with $N$ fundamental and $N$ anti-fundamental hypermultiplets. the index reads
\begin{align}
	I = \oint_{|z_i| = 1} \prod_{A = 1}^{N - 1} \frac{dz_A}{2\pi i z_A} Z_\text{1-loop-VM}(z) Z_\text{1-loop-HM}(z) |Z_\text{inst}(Q; z, \mu^\epsilon, \tilde \mu^\epsilon; \mathfrak{p}, \mathfrak{q})|^2 \ ,
\end{align}
where (with $\prod_{A = 1}^N z_A = 1$ imposed implicitly everywhere)
\begin{align}
	Z_\text{1-loop-VM} \equiv & \ \frac{(\mathfrak{p}; \mathfrak{p}, \mathfrak{q})^{N-1}(\mathfrak{q}; \mathfrak{p}, \mathfrak{q})^{N-1}}{N!}\prod_{\substack{A, B = 1\\A\ne B}}^N
	(z_Az_B^{-1}; \mathfrak{p},\mathfrak{q})
	(z_Az_B^{-1}\mathfrak{pq};\mathfrak{p},\mathfrak{q}) \ ,\\
	Z_\text{1-loop-HM} \equiv & \ \prod_{A = 1}^N \prod_{i = 1}^N \frac{1}{(\sqrt{\mathfrak{pq}} z_A b^{-1}\mu_i^{-1};\mathfrak{p},\mathfrak{q}) (\sqrt{\mathfrak{pq}} z^{-1}_A b\mu_i ;\mathfrak{p},\mathfrak{q})}\frac{1}{(\sqrt{\mathfrak{pq}} z_A \tilde \mu_i^{-1};\mathfrak{p},\mathfrak{q}) (\sqrt{\mathfrak{pq}} z^{-1}_A \tilde \mu_i;\mathfrak{p},\mathfrak{q})}  \ .\nonumber
\end{align}
Here we have separate the $U(1)$ flavor fugacity $b$ from the $SU(N)$ flavor fugacities $\mu_i$ satisfies $\prod_{i = 1}^{N} \mu_i = 1$. $\mathfrak{p}$ and $\mathfrak{q}$ are fugacites associated to the $U(1) \times U(1)$ rotations of $S^4$, related to the exponentiated $\Omega$-deformation parameters $\epsilon_{1,2}$ in 4d by $\mathfrak{p} \equiv e^{2\pi i \epsilon_1}$, $\mathfrak{q} \equiv e^{2\pi i \epsilon_2}$. We choose $|\mathfrak{p}|, |\mathfrak{q}| < 1$ as usual.

The baryonic simple poles that we will be interested in are
\begin{align}
	b \to \mathfrak{p}^{n^\text{L} + \frac{1}{2}} \mathfrak{q}^{n^\text{R} + \frac{1}{2}} \ , \qquad n^\text{L}, n^\text{R} \in \mathbb{N} \ .
\end{align}
At these values, the poles of the following form, coming from the fundamental hypermultiplet one-loop contributions, of the integrand pinch the contour since there are only $N-1$ independent variables $z_A$,
\begin{align}
	z_A = b \mu_{\iota_A} \mathfrak{p}^{ - n^\text{L}_A - \frac{1}{2}} \mathfrak{q}^{- n^\text{R} - \frac{1}{2}} \ , \qquad \sum_{A = 1}^N n_A^\text{L,R} = n^\text{L,R} \ , \qquad \iota \in S_N \ ,
\end{align}
and as a result, the contour integral picks up residues at these poles, which is a sum over integer partitions of $n^\text{L}$ and $n^\text{R}$ (on top of the sum over Higgs vacua $\iota$). The residue can also be extracted by performing a contour integral of $b$, which effectively gauges $U(1)$ flavor symmetry and making the $SU(N)$ theory into a $U(N)$ theory. By a redefinition of variables, $b$ can be absorbed into the $z$'s making all $N$ variables $z_A$ independent, and one simply needs to collect the residues of the integrand at
\begin{align}
	z_A = \mu_{\iota_A} \mathfrak{p}^{- n_A^\text{L} - \frac{1}{2}}\mathfrak{q}^{- n_A^\text{R} - \frac{1}{2}} \ .
\end{align}
A symmetry observation implies that the dependence of the residue on the Higgs vacua is trivial, and one may simply put $\iota_A = A$, the degeneracy then removes the $\frac{1}{N!}$ up front.

\begin{figure}
	\centering
	\includegraphics[width=0.8\textwidth]{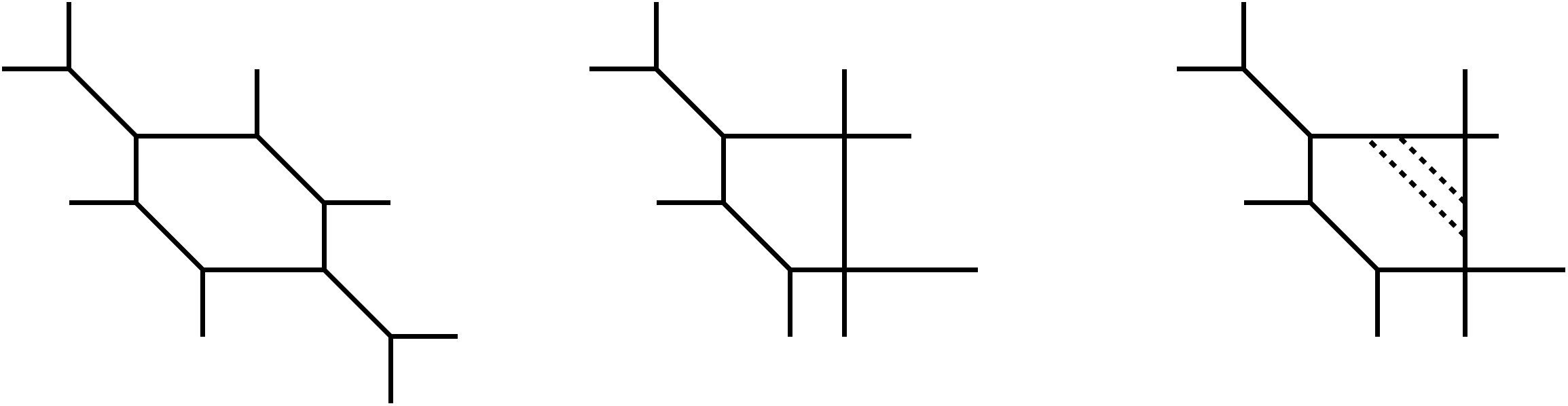}
	\caption{\label{pq-web-Higgsing}The figure on the left is a $(p,q)$-web diagram that engineers a $SU(2)$ SQCD, which can be constructed by gluing two strip geometry. In the middle figure, the Coulomb branch parameter has been tuned to special values corresponding to the root of the Higgs branch, where some NS5 branes and D5-branes are aligned reconnected. In the figure on the right, the NS5-brane on the right is pulled away of the web while suspending some D3-branes (represented by the dashed lines) which engineer codimension-two defects in the remaining 5d theory, which is a free theory with $4$ fundamentals in this case.}
\end{figure}

One may look at the brane construction of such procedure. Consider a linear $SU(N)$ quiver gauge theory theory $\mathcal{T}$ engineered by a five-brane web in type IIB string theory; see figure \ref{pq-web-Higgsing} for an simplest example. We also tabulate here the spacetime directions spanned by the branes.
\begin{center}
	\begin{tabular}{c c c c c c c c c c c}
		& 0 &  1 & 2 & 3 & 4 & 5 & 6 & 7 & 8 & 9\\
		\hline
		D5  & $-$ & $-$ & $-$ & $-$ & $-$ & $\cdot$ & $-$ & $\cdot$ & $\cdot$ & $\cdot$ \\
		NS5 & $-$ & $-$ & $-$ & $-$ & $-$ & $-$ & $\cdot$ & $\cdot$ & $\cdot$ & $\cdot$ \\
		(1,1) & $-$ & $-$ & $-$ & $-$ & $-$ & / & / & $\cdot$ & $\cdot$ & $\cdot$ \\
		D3$^\text{L}$ & $-$ & $-$ & $-$ & $\cdot$ & $\cdot$ & $\cdot$ & $\cdot$ & $-$ & $\cdot$ & $\cdot$ \\
		D3$^\text{R}$ & $-$ & $\cdot$ & $\cdot$ & $-$ & $-$ & $\cdot$ & $\cdot$ & $-$ & $\cdot$ & $\cdot$ 
	\end{tabular}
\end{center}
The horizontal external legs on the two sides represents fundamental/anti-fundamental hypermultiplets associated to the manifest $SU(N) \times SU(N)$ flavor symmetry. Gauging in additional $N^2$ hypermultiplets to one side corresponds to gluing in an additional strip of D5-NS5 geometry to that side of the web, making a $SU(N)$ quiver gauge theory with one additional gauge node. One can then tune the adjacent Coulomb branch parameter to the root of Higgs branch, or, in terms of the $(p,q)$ web, by aligning the D5/NS5 branes on the side. One can further pull away the NS5 brane from the web while stretching $n^\text{L}$ D3$^\text{L}$ and $n^\text{R}$ D3$^\text{R}$ branes which extend in different directions as shown in the above table. These D3-branes support the worldvolume theories of the codimension-two defects in the bulk 5d theory.

The $(p,q)$ web diagram can also be understood in terms of refined topological string, which straightforwardly produces the Nekrasov partition function of the 5d theory. In particular, the S-duality of the $(p,q)$-web leaves the partition function invariant, which is associated to identities involving the (skew-) Macdonald polynomials.

\section{Intersecting surface defects on \texorpdfstring{$S^4 \times S^1$}{S4xS1}}

\subsection{Higgsing}

Let us consider $\mathcal{N} = 1$ $U(N)$ gauge theories coupled with $N$ fundamental, $N$ anti-fundamental hypermultiplet on $S^4 \times S^1$. The index reads \cite{Gaiotto:2014ina,Kim:2012gu,Kim:2013nva},
\begin{align}
	I = \oint \prod_{A = 1}^N \frac{dz_A}{2\pi i z_A} Z_\text{1-loop-VM}(z) Z_\text{1-loop-HM}(z) |Z_\text{inst}(Q; z, \mu^\epsilon, \tilde \mu^\epsilon; \mathfrak{p}, \mathfrak{q})|^2 \ ,
\end{align}
For convenience, we reproduce the one-loop factors here,
\begin{align}
	Z_\text{1-loop-VM} \equiv & \ \frac{1}{N!}\prod_{\substack{A, B = 1\\A\ne B}}^N
	(z_Az_B^{-1}; \mathfrak{p},\mathfrak{q})
	(z_Az_B^{-1}\mathfrak{pq};\mathfrak{p},\mathfrak{q}) \ ,\\
	Z_\text{1-loop-HM} \equiv & \ \prod_{A = 1}^N \prod_{i = 1}^N \frac{1}{(\sqrt{\mathfrak{pq}} z_A \mu_i^{-1};\mathfrak{p},\mathfrak{q}) (\sqrt{\mathfrak{pq}} z^{-1}_A \mu_i;\mathfrak{p},\mathfrak{q})}\frac{1}{(\sqrt{\mathfrak{pq}} z_A \tilde \mu_i^{-1};\mathfrak{p},\mathfrak{q}) (\sqrt{\mathfrak{pq}} z^{-1}_A \tilde \mu_i;\mathfrak{p},\mathfrak{q})}  \ .
\end{align}
Here $\mu$ and $\tilde \mu$ are flavor fugacities for the fundamental and anti-fundamental hypermultiplets\footnote{The distinguishing of anti-fundamental from fundamental hypermultiplets is merely for bookkeeping purpose.}.

The instanton partition function is given by a sum over Young diagrams
\begin{align}
	Z_\text{inst}(Q; z, \mu^\epsilon, \tilde \mu^\epsilon; \mathfrak{p}, \mathfrak{q})
	= \sum_{\vec Y} Q^{|\vec Y|} \mathcal{Z}_\text{CS}(z)\mathcal{Z}_\text{VM}(z)\mathcal{Z}_\text{HM}(z, \mu^\epsilon, \tilde \mu^\epsilon) \ ,
\end{align}
where relevant factors are collected in appendix \ref{app:ipf}, and $Q$ encodes the Yang-Mills coupling constant. The superscript $\epsilon$ denotes the shift
\begin{align}
	\mu^\epsilon = \mu \sqrt{\mathfrak{pq}}, \qquad \tilde \mu^\epsilon = \tilde \mu \sqrt{\mathfrak{pq}} \ .
\end{align}
The modular square is defined by inverting $Q$ together with all other fugacities,
\begin{align}
	|f(Q; z, \mu, \tilde \mu; \mathfrak{p}, \mathfrak{q})|^2
	= f(Q; z, \mu, \tilde \mu; \mathfrak{p}, \mathfrak{q}) f(Q^{-1}; z^{-1}, \mu^{-1}, \tilde \mu^{-1}; \mathfrak{p}^{-1}, \mathfrak{q}^{-1})
	 \ .
\end{align}

To access the partition functions of intersecting surface defects, we focus on the following set of simple poles labeled by two partitions $\{n^\text{L,R}_A\}$ outside the unit circles of $z_A$,
\begin{align}
	z_A = \mu_A \mathfrak{p}^{- n_A^\text{L} - \frac{1}{2}}\mathfrak{q}^{- n_A^\text{R} - \frac{1}{2}}\ , \qquad
	\sum_A n_A^\text{L} = n^\text{L},  \qquad
	\sum_A n_A^\text{R} = n^\text{R}, 
\end{align}
where $n^\text{L, R} \in \mathbb{N}$ are fixed natural numbers. These poles come from the perturbative one-loop factors associated with the fundamental hypermultiplets, while the remaining perturbative and non-perturbative factors are regular at these values.

The residue of the one-loop factors can be computed straightforwardly and simplified using the shift properties of the double $q$-Pochhammer symbol. The anti-fundamental factor evaluated at the pole gives
\begin{align}
	Z_\text{1-loop-afund}\Big|_\text{pole} = Z^{S^4\times S^1}_\text{free} Z^{\vec n^\text{L}}_\text{afund}Z^{\vec n^\text{R}}_\text{afund} Z_\text{afund-extra} \ ,
\end{align}
where $Z_\text{free}$ is the 5d index of $N^2$-free hypermultiplets with flavor fugacity $\mu_i \tilde \mu_j^{-1}$,
\begin{align}
	Z_\text{free}^{S^4 \times S^1} = \prod_{I,J = 1}^{N} \frac{1}{(\sqrt{\mathfrak{pq}}\mu_I \tilde \mu_J^{-1}; \mathfrak{p}, \mathfrak{q})(\sqrt{\mathfrak{pq}}\mu^{-1}_I \tilde \mu_J; \mathfrak{p}, \mathfrak{q})} \ .
\end{align}
We shall come back to the remaining factors momentarily. The residue of vector multiplet and fundamental hypermultiplets factors read
\begin{align}
	Z_\text{1-loop-VM} \operatorname{Res}Z_\text{1-loop-fund}\Big|_\text{pole}
	= 
	Z^{\vec n^\text{L}}_\text{fund+adj}Z^{\vec n^\text{R}}_\text{fund+adj} Z_\text{VF-extra} \ .
\end{align}
Here and in the above, $Z^{\vec n^\text{L,R}}_\text{matter}$ denotes the one-loop factor from the said matter in the Higgs-branch-localized form of the 3d $\mathcal{N} = 2$ index of the world-volume gauge theory living on $S^2_\text{L,R} \times S^1$. The factor $Z_\text{matter-extra}$ are simple factors that depend on $n^\text{L,R}_A$ which soon participate in partial cancellation with similar factors from the evaluation of the instanton partition function.

Next we turn to the instanton partition function which is a sum over tuples $\{Y_A\}_{A = 1}^N$ of Young diagrams. It can be seen immediately that the sum is truncated to a sum over tuples of hook Young diagrams $\{Y_A\}_{A = 1}^N$ where each $Y_A$ does not contain the ``forbidden box'' at $(n_A^\text{L} + 1, n_A^\text{R} + 1)$. This is because the fundamental hypermultiplets contribute a factor
\begin{align}
	& \ \prod_{I = 1}^{N}\prod_{A = 1}^{N}\prod_{(r, s) \in Y_A} 2\sinh \pi i \beta (\hat z_A - \hat \mu^\epsilon_I + r \epsilon_1 + s \epsilon_2)\\
	\to & \ \prod_{I = 1}^{N}\prod_{A = 1}^{N}\prod_{(r, s) \in Y_A}\sinh \pi i \beta \left[(\hat z_A - \hat \mu_I) + (r - n^\text{L}_A - 1) \epsilon_1 + (s  - n^\text{R}_A - 1) \epsilon_2\right] \ ,
\end{align}
which vanishes whenever one $Y_A$ contains that box. As explained in \cite{Gomis:2016ljm,Pan:2016fbl}, it is crucial to divide these hook Young diagrams into two classes, ``large'' and ``small'', depending on whether the diagram contains or does not contain the $n_A^\text{L} \times n_A^\text{R}$ rectangle of boxes containing the box $(1,1)$. It is easiest to deal with the contributions from the large Young diagrams. First of all, each large Yong diagram can be alternatively described by two subdiagrams located at the lower left and upper right corner, sandwiching the rectangle. The lower left subdiagram will be called $Y^\text{L}_A$ while the one at the upper right $Y^\text{R}_A$. From them, one can further define two non-decreasing sequences of natural numbers
\begin{figure}
	\centering
	\includegraphics{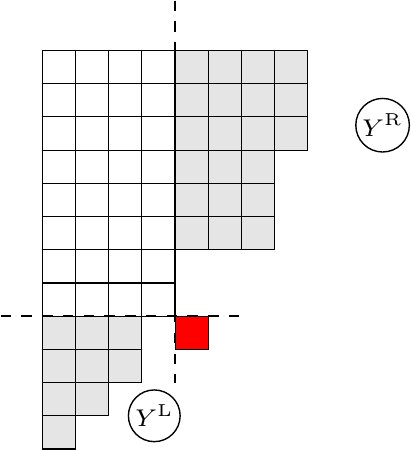}
	~~~~~~~~
	\includegraphics{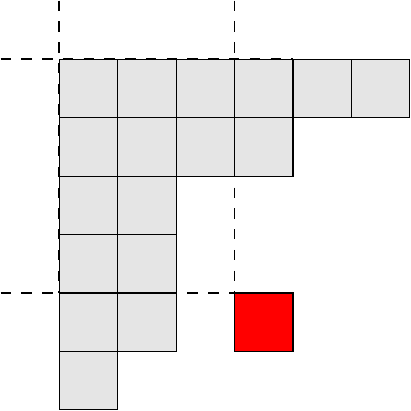}
	\caption{In the left figure, a large Young diagram $Y$ is depicted with a red ``forbidden box'', where we identify the subdiagrams $Y^\text{L,R}$ as the shaded parts. Apparently, the large diagram contains a $n^\text{L} \times n^\text{R}$ (width $\times$ height)-rectangle of boxes colored in white. On the right, we depict a small Young diagram.}
\end{figure}
\begin{align}
	\mathfrak{m}^\text{L}_{A\mu} \equiv Y^\text{L}_{A,n^\text{L}_A - \mu}, \quad \mu = 0, \ldots, n_A^\text{L} - 1 \ ,
	\qquad \mathfrak{m}^\text{R}_{A\nu} = (Y^\text{R})^\vee_{A,n^\text{R}_A - \nu} \ , \quad \nu = 0, \ldots, n_A^\text{R} - 1\ ,
\end{align}
where $\vee$ denotes transposition of a Young diagram. The contributions from such a tuple of large Young diagram at the pole factorizes into
\begin{align}
	\text{large diagram} \to Z_\text{vortex}^{\vec n^\text{L}}(\mathfrak{m};k_\text{CS}^\text{5d},Q; t, \tilde t, v, \tau, q)_\text{L}
	Z_\text{vortex}^{\vec n^\text{R}}(\mathfrak{m};k_\text{CS}^\text{5d},Q; t, \tilde t, v, \tau, q)_\text{R}\ ,\\
	\times Z_\text{intersection}^{\vec n^\text{L}, \vec n^\text{R}} (\mathfrak{m}^\text{L}, \mathfrak{m}^\text{R})
	Z_\text{cl-extra}^{\vec n^\text{L}, \vec n^\text{R}}
	(Z_\text{afund-extra}^{\vec n^\text{L}, \vec n^\text{R}}Z_\text{VF-extra}^{\vec n^\text{L}, \vec n^\text{R}})^{-1}\ ,
\end{align}
where (summands of) vortex partition functions on $S^2 \times_{q_\text{L,R}} S^1_\beta$ appears with 3d fugacities 
\begin{align}\label{5d-3d-fugacities}
	& (t_i \tau)_\text{L} = \mu_i \mathfrak{p}^{-1}\mathfrak{q}^{ - 1/2} \ ,\qquad
	(\tilde t_i \tau)_\text{L} = \tilde \mu_i^{-1}\mathfrak{q}^{1/2} \ ,\qquad
	v_\text{L} = \mathfrak{p}^{-1}, \qquad q_\text{L} = \mathfrak{q}\ , \\
	& (t_i \tau)_\text{R} = \mu_i \mathfrak{q}^{-1}\mathfrak{p}^{ - 1/2} \ ,\qquad
	(\tilde t_i \tau)_\text{R} = \tilde \mu_i^{-1}\mathfrak{p}^{1/2} \ ,\qquad
	v_\text{R} = \mathfrak{q}^{-1}, \qquad q_\text{R} = \mathfrak{p} \ .
\end{align}
It is easy to observe the fugacity relations
\begin{align}
	(t_i\tau \mathfrak{q}^{-1/2})_\text{L} = (t_i\tau \mathfrak{q}^{-1/2})_\text{R} \ ,\qquad
	(\tilde t_i\tau \mathfrak{q}^{-1/2})_\text{L} = (\tilde t_i\tau \mathfrak{q}^{-1/2})_\text{R}\ ,
\end{align}
which arise from superpotentials coupling the free 5d hypermultiplets and the 3d chiral multiplets. The factors $(Z_\text{afund-extra}^{\vec n^\text{L}, \vec n^\text{R}}Z_\text{VF-extra}^{\vec n^\text{L}, \vec n^\text{R}})^{-1}$ appearing in the factorization cancel those from the one-loop factors when computing the residue. The classical extra factor reads
\begin{align}
	Z^{\vec n^\text{L}, \vec n^\text{R}}_\text{cl-extra} =
	\prod_{A = 1}^{N}
	Q^{n^\text{L}_A n^\text{R}_A}
	(\mu
   \mathfrak{p}^{-\frac{1}{2} (n^\text{L}_A + 1)}
	 \mathfrak{q}^{-\frac{1}{2} (n^\text{R}_A + 1)})^{- k_\text{CS}^\text{5d}n^\text{L}_A n^\text{R}_A} \ .
\end{align}
Obviously, such factor is independent of $\vec Y$ can be relocated outside of the sum over hook Young diagrams. Finally the intersection factor is the most crucial factor in the following discussion, which reads
\begin{align}
	Z_\text{intersection}^{\vec n^\text{L}, \vec n^\text{R}}
	= & \ \prod_{A, B}^{N}\prod_{\mu = 0}^{n^\text{L}_A - 1} \prod_{\nu = 0}^{n^\text{R}_B - 1}\frac{1}{2 \sinh \pi i \beta (\hat \mu_{AB} + (\mathfrak{m}^\text{L}_{A\mu} + \nu) \epsilon_2 - (\mathfrak{m}^\text{R}_{B\nu} + \mu)\epsilon_1 - \epsilon_1)}\\
	& \ \times \prod_{A, B}^{N}\prod_{\mu = 0}^{n^\text{L}_A - 1} \prod_{\nu = 0}^{n^\text{R}_B - 1}\frac{1}{2 \sinh \pi i \beta (\hat \mu_{AB} + (\mathfrak{m}^\text{L}_{A\mu} + \nu) \epsilon_2 - (\mathfrak{m}^\text{R}_{B\nu} + \mu)\epsilon_1 + \epsilon_2)} \ .
\end{align}
One can view this factor as a product over the boxes inside the $n^\text{L}_A \times n^\text{R}_A$ rectangle region, which of course precisely fill the entire region since we are dealing with large Young diagrams.

The contributions from the small Young diagram tuples are less trivial. Nonetheless, one can still define non-decreasing integers (which however could be negative for $\mathfrak{m}^\text{L}$) by
\begin{align}
	\mathfrak{m}^\text{L}_{A\mu} \equiv Y_{A,n^\text{L}_A - \mu} - n^\text{L}_A, \quad \mu = 0, \ldots, n_A^\text{L} - 1 \ ,
	\qquad \mathfrak{m}^\text{R}_{A\nu} = (Y^\text{R})^\vee_{A,n^\text{R}_A - \nu} \ , \quad \nu = 0, \ldots, n_A^\text{R} - 1\ .
\end{align}
Note that by definition, for a small Young diagram $Y^\text{L}_{A, r} \le n_A^\text{L}$, $r = 1, \ldots, n_A^\text{L}$. The contribution from a tuple of small (or a tuple containing both small and large) Young diagram reads
\begin{align}
	\text{small diagram} \to Z_\text{semi-vortex}^{\vec n^\text{L}}(\mathfrak{m};k_\text{CS}^\text{5d},Q; t, & \tilde t, v, \tau, q)_\text{L}
	Z_\text{vortex}^{\vec n^\text{R}}(\mathfrak{m};k_\text{CS}^\text{5d},Q; t, \tilde t, v, \tau, q)_\text{R} \\
	& \times {Z'}_\text{intersection}^{\vec n^\text{L}, \vec n^\text{R}} (\mathfrak{m}^\text{L}, \mathfrak{m}^\text{R})
	Z_\text{cl-extra}^{\vec n^\text{L}, \vec n^\text{R}}
	(Z_\text{afund-extra}^{\vec n^\text{L}, \vec n^\text{R}}Z_\text{VF-extra}^{\vec n^\text{L}, \vec n^\text{R}})^{-1}\ . \nonumber
\end{align}
Here, as indicated by the prime, the intersection factor ${Z'}_\text{intersection}^{\vec n^\text{L}, \vec n^\text{R}}$ is a product of the same factors as in the large Young diagram case, but now only over the boxes inside the rectangle region.

Putting both the large and small Young diagram contributions together, we have the instanton partition function evaluated at the pole
\begin{align}
	\Bigg|
	Z_\text{cl-extra}^{\vec n^\text{L}, \vec n^\text{R}}
	Z_\text{afund-extra}^{\vec n^\text{L}, \vec n^\text{R}}
	Z_\text{fund-extra}^{\vec n^\text{L}, \vec n^\text{R}}
	\bigg[
	  \sum_\text{large} & \ Z_\text{vortex}^{\vec n^\text{L}}(\mathfrak{m}^\text{L})
	Z_\text{vortex}^{\vec n^\text{R}}(\mathfrak{m}^\text{R}) Z_\text{intersection}^{\vec n^\text{L}, \vec n^\text{R}}(\mathfrak{m}^\text{L}, \mathfrak{m}^\text{R})\\
	& \ + \sum_\text{small} Z_\text{semi-vortex}^{\vec n^\text{L}}(\mathfrak{m}^\text{L})
	Z_\text{vortex}^{\vec n^\text{R}}(\mathfrak{m}^\text{R}) Z_\text{intersection}^{\vec n^\text{L}, \vec n^\text{R}}(\mathfrak{m}^\text{L}, \mathfrak{m}^\text{R})
	\bigg]
	\Bigg|^2 \ . \nonumber
\end{align}
To avoid clutter, we have omitted the fugacities from the expression. Immediately we recall that the modular-square is defined by inverting all fugacities including $Q$, and since the $Z_\text{cl-extra}$ is a monomial of these fugacity, the classical extra factor actually cancel within the $| \ldots |^2$. As advertise before, the extra factors from the anti-fundamental and fundamental hypermultiplets will annihilate with those from the residue computation of the perturbative one-loop factor. Altogether, we have the sum of residues given by
\begin{align}
	= Z^{S^4\times S^1}_\text{free} 
	\sum_{\vec n^\text{L}, \vec n^\text{R}} 
	& Z^{\vec n^\text{L}}_\text{afund}Z^{\vec n^\text{R}}_\text{afund} Z^{\vec n^\text{L}}_\text{fund+adj}Z^{\vec n^\text{R}}_\text{fund+adj}\nonumber\\
	& \ \times \Bigg|
	\bigg[
	  \sum_\text{large} Z_\text{vortex}^{\vec n^\text{L}}(\mathfrak{m}^\text{L})
	Z_\text{vortex}^{\vec n^\text{R}}(\mathfrak{m}^\text{R}) Z_\text{intersection}^{\vec n^\text{L}, \vec n^\text{R}}(\mathfrak{m}^\text{L}, \mathfrak{m}^\text{R})\\
	& \ \qquad \quad + \sum_\text{small} Z_\text{semi-vortex}^{\vec n^\text{L}}(\mathfrak{m}^\text{L})
	Z_\text{vortex}^{\vec n^\text{R}}(\mathfrak{m}^\text{R}) Z_\text{intersection}^{\vec n^\text{L}, \vec n^\text{R}}(\mathfrak{m}^\text{L}, \mathfrak{m}^\text{R})
	\bigg]
	\Bigg|^2 \ .\nonumber
\end{align}
We will see shortly that this expression can be reorganized into the index of a 5d/3d/1d coupled system.

\subsection{Index of intersecting gauge theory}

We are now ready to identify the above result from Higgsing the 5d $\mathcal{N} = 1$ theory with the index of a free 5d theory coupled to a $U(n^\text{L}) \times  U(n^\text{R}) $ gauge theory on an intersecting space $(S^2_\text{L}\times_{q_\text{L}} S^1) \cup (S^2_\text{R}\times_{q_\text{R}} S^1)$. The $\Omega$-deformation parameters are given by $q_\text{L} = e^{2\pi i \beta \epsilon_2}$, $q_\text{R} = e^{2\pi i \beta \epsilon_1}$. The gauge theory is coupled to two copies of $n_\text{f} = n_\text{af} = N$ fundamental and ant-fundamental chiral multiplets and one adjoint chiral multiplets living on two different $S^2_\text{L} \times_{q_\text{L}} S^1$ and $S^2_\text{R} \times_{q_\text{R}} S^1$, and additionally a pair of bifundamental 1d $\mathcal{N} = 2$ chiral multiplet living at each of the two common circle intersection $\{N\} \times S^1$ and $\{S\} \times S^1$, where $N$ and $S$ refer to the common north and south poles of the two $S^2$'s.
\begin{figure}
	\centering
	\includegraphics{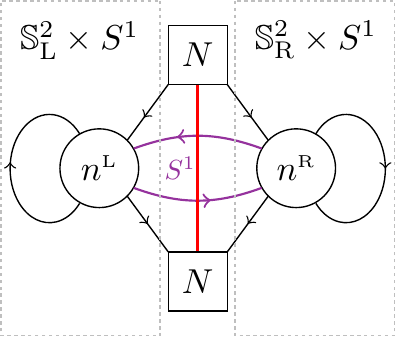}
	\caption{\label{intersecting-SQCDA}
	The worldvolume theory of the surface defect is described by the quiver diagram. Each of the two unitary gauge theories on each $S^2 \times S^1$ is coupled to $N$ fundamental, $N$ anti-fundamental and one adjoint chiral multiplets, and further to the 1d bifundamental chiral multiplet supported on the intersecting, as colored in purple. The red line above indicates the free theory of $N^2$ hypermultiplets on $S^4 \times S^1$.
	}
\end{figure}
We propose the index to be given simply by the sum of contour integrals
\begin{align}
	& \ I^{n^\text{L}, n^\text{R}} \\
	= & \ \sum_{\vec B^\text{L}, \vec B^\text{R}}\oint
	\prod_{a = 1}^{n^\text{L}} \frac{dz^\text{L}_a}{2\pi i z^\text{L}_a}
	\prod_{a = 1}^{n^\text{R}} \frac{dz^\text{R}_a}{2\pi i z^\text{R}_a}
	Z^{S^2_\text{L}\times_{q_\text{L}} S^1}(z^\text{L}, B^\text{L})
	Z^{S^2_\text{L}\times_{q_\text{R}} S^1}(z^\text{R}, B^\text{R})
	Z^{S^1}_\text{intersection}(z^\text{L}, B^\text{L}, z^\text{R},B^\text{L}) \ ,\nonumber
\end{align}
where $Z^{S^2_\text{L,R}\times_{q_\text{L,R}} S^1}(z^\text{L,R})$ denotes the usual integrand of the index of a $U(n^\text{L,R})$ SQCDA with $n_\text{f} = n_\text{af} = N$, while the last factor $Z^{S^1}_\text{intersection}(z^\text{L}, z^\text{R})$ captures the contribution from the one-dimensional bifundamental chiral multiplet,
\begin{align}
	Z^{S^1}_\text{intersection}(z^\text{L}, z^\text{R})
	= & \ \prod_{a = 1}^{n^\text{L}}\prod_{b = 1}^{n^\text{R}} \frac{1}{2\sinh \pi i \beta ((\hat z^\text{L}_a + \frac{1}{2}B_a^\text{L}\epsilon_2) - (\hat z^\text{R}_b + \frac{1}{2}B^\text{R}_a\epsilon_1) \pm \frac{1}{2}(\epsilon_1 + \epsilon_2))} \nonumber\\
	& \ \times \prod_{a = 1}^{n^\text{L}}\prod_{b = 1}^{n^\text{R}} \frac{1}{2\sinh \pi i \beta ((\hat z^\text{L}_a - \frac{1}{2}B_a^\text{L}\epsilon_2) - (\hat z^\text{R}_b - \frac{1}{2}B^\text{R}_a\epsilon_1) \pm \frac{1}{2}(\epsilon_1 + \epsilon_2))}  \ .
\end{align}
Here the first line comes from the degrees of freedom localized at the north pole $\{N\} \times S^1$ and the second lines at the south pole $\{S\} \times S^1$. The quiver diagram of the theory associated to this superconformal index is given in Figure \ref{intersecting-SQCDA}.

To identify the index with the result from the previous section, it is crucial to correctly specify the integration contour. We require that the contour integral picks up residues of four types of poles, which we call \textit{type old}, \textit{type N}, \textit{type S} and \textit{type NS} \cite{Pan:2016fbl,Gomis:2016ljm,1993alg.geom..7001J}, all of which depend on a choice of integer partitions $\{n^\text{L,R}_i \}_{i = 1}^N$ labeling the isolated massive Higgs branch vacua. Doing so, the color indices $a/b = 1, \ldots, n^\text{L}/n^\text{R}$ will be reorganized into $\{(i\mu)|i = 1,\ldots, N, \mu = 0, \ldots, n^\text{L,R}_i - 1\}$.

The poles of type old are simply the familiar poles of the factors $Z^{S^2_\text{L,R} \times S^1}$ that one would pick up to perform factorization of the usual index on $S^2_\text{L,R} \times S^1$ separately. Concretely,
\begin{align}
	\text{type old: } \qquad (z_{i\mu}q^{\frac{B}{a}})_\text{L,R} = (t_i \tau v^{\mu} q^{\mathfrak{m}_{i \mu}})_\text{L,R},
	\qquad
	(z_{i\mu}q^{ - \frac{B}{a}})_\text{L,R} = (t_i \tau v^{\mu} q^{\overline{\mathfrak{m}}_{i \mu}})_\text{L,R} \ ,
\end{align}
where $\mathfrak{m}^\text{L,R}_{i \mu}, \overline{\mathfrak{m}}^\text{L,R}_{i\mu}$ are non-decreasing sequences of natural numbers, e.g., $0 \le\mathfrak{m}_{i \mu} \le \mathfrak{m}_{i, \mu + 1}$. The other three types of poles arise from first taking a subset of the standard $z^\text{R}$-poles from $Z^{S^2_\text{R}\times S^1}$, and then a combination of $z^\text{L}$-poles from $Z^{S^2_\text{L} \times S^1}$ and the intersection factor. Concretely, they are poles of the form
\begin{align}
	(z_{i\mu}q^{\frac{B}{a}})_\text{L,R} = (t_i \tau v^{\mu} q^{\mathfrak{m}_{i \mu}})_\text{L,R},
	\qquad
	(z_{i\mu}q^{ - \frac{B}{a}})_\text{L,R} = (t_i \tau v^{\mu} q^{\overline{\mathfrak{m}}_{i \mu}})_\text{L,R}\ ,
\end{align}
where the non-decreasing sequences $\mathfrak{m}^\text{L}, \overline{\mathfrak{m}}^\text{L}$ take values in some different range than natural numbers, determined for each type by a set of integers $\hat \nu_i \in \{-1, 0, \ldots, n^\text{R} - 1\}$ where not all $\hat \nu_i = -1$:
\begin{itemize}
	\item Type N$_{\hat \nu}$:
	\begin{align}
		& \mathfrak{m}^\text{R}_{i, n^\text{R}_i - 1} \ge \ldots \ge \mathfrak{m}^\text{R}_{i \hat \nu_i} = \ldots = \mathfrak{m}^\text{R}_{i0} = 0\ ,& &
		\overline{\mathfrak{m}}^\text{R}_{i, n^\text{R} - 1} \ge \ldots \ge \overline{\mathfrak{m}}^\text{R}_{i0} \ge 0 \nonumber \\
		& \mathfrak{m}^\text{L}_{i, n^\text{L}_i - 1} \ge \ldots \ge \mathfrak{m}_{i1}^\text{L} \ge \mathfrak{m}^\text{L}_{i 0} \left\{
		\begin{matrix}
			= - (\hat \nu_i + 1) \ ,&  \nu_i \ge 0 \\
			\ge 0 \ , & \nu_i = -1 \hfill
		\end{matrix}
		\right . \ ,& &
		\overline{\mathfrak{m}}^\text{L}_{i, n^\text{L} - 1} \ge \ldots \ge \overline{\mathfrak{m}}^\text{L}_{i0} \ge 0 \ .
	\end{align}
  \item Type S$_{\hat \nu}$:
  \begin{align}
  	& \mathfrak{m}^\text{R}_{i, n^\text{R} - 1} \ge \ldots \ge \mathfrak{m}^\text{R}_{i0} \ge 0\ ,& &
  	\overline{\mathfrak{m}}^\text{R}_{i, n^\text{R}_i - 1} \ge \ldots \ge \overline{\mathfrak{m}}^\text{R}_{i \hat \nu_i} = \ldots = \overline{\mathfrak{m}}^\text{R}_{i0} = 0 \nonumber
  	 \\
  	& \mathfrak{m}^\text{L}_{i, n^\text{L} - 1} \ge \ldots \ge \mathfrak{m}^\text{L}_{i0} \ge 0 \ ,& &
  	\overline{\mathfrak{m}}^\text{L}_{i, n^\text{L}_i - 1} \ge \ldots \ge \overline{\mathfrak{m}}_{i1}^\text{L} \ge \overline{\mathfrak{m}}^\text{L}_{i 0} \left\{
		\begin{matrix}
			= - (\hat \nu_i + 1) \ ,&  \nu_i \ge 0 \\
			\ge 0 \ , & \nu_i = -1 \hfill
		\end{matrix}
		\right .\ .
  \end{align}
  \item Type NS$_{\hat \nu^\text{N} \hat \nu^\text{S}}$:
    \begin{align}
    	& \mathfrak{m}^\text{R}_{i, n^\text{R}_i - 1} \ge \ldots \ge \mathfrak{m}^\text{R}_{i \hat \nu^\text{N}_i} = \ldots = \mathfrak{m}^\text{R}_{i0} = 0\ ,
    	\qquad
    	\overline{\mathfrak{m}}^\text{R}_{i, n^\text{R}_i - 1} \ge \ldots \ge \overline{\mathfrak{m}}^\text{R}_{i \hat \nu^\text{S}_i} = \ldots = \overline{\mathfrak{m}}^\text{R}_{i0} = 0 \ ,\nonumber
    	 \\
    	& \mathfrak{m}^\text{L}_{i, n^\text{L}_i - 1} \ge \ldots \ge \mathfrak{m}_{i1}^\text{L} \ge \mathfrak{m}^\text{L}_{i 0} \left\{
  		\begin{matrix}
  			= - (\hat \nu^\text{N}_i + 1) \ ,&  \nu^\text{N}_i \ge 0 \\
  			\ge 0 \ , & \nu^\text{N}_i = -1 \hfill
  		\end{matrix}
  		\right .\ , \\
    	& \overline{\mathfrak{m}}^\text{L}_{i, n^\text{L}_i - 1} \ge \ldots \ge \overline{\mathfrak{m}}_{i1}^\text{L} \ge \overline{\mathfrak{m}}^\text{L}_{i 0} \left\{
  		\begin{matrix}
  			= - (\hat \nu^\text{S}_i + 1) \ ,&  \nu^\text{S}_i \ge 0 \\
  			\ge 0 \ , & \nu^\text{S}_i = -1 \hfill
  		\end{matrix}
  		\right .\ . \nonumber
    \end{align}
\end{itemize}
Obviously, when all $\hat \nu_i = -1$ one would recover poles of type old which we have avoided in the definition.

It is straightforward, though somewhat tedious, to check that summing the residues from all four types of poles recovers the result from Higgsing. Indeed, the poles of type old are in one-to-one correspondence with the double sum over tuples of large Young diagrams in the instanton part, which is a product of the north and south pole contributions. Poles of type N then correspond to the double sum of tuples of large and small diagrams in the north and south pole respectively, while poles of type S correspond to the other way around. Finally, poles of type NS correspond to the double sum over tuples of small Young diagrams.


\section{3d Dualities}

\subsection{3d Mirror symmetry}

It is well-known that various 3d $\mathcal{N} = 4$ theories enjoy an IR duality called 3d mirror symmetry, where two UV supersymmetric gauge theories flow in the IR to two SCFTs with the Higgs branch of one SCFT identified with the Coulomb branch of the other, among many other identifications of physical quantities. Such duality can usually be realized by the S-duality in type IIB string theory acting on brane systems that engineer these gauge theories. 3d $\mathcal{N} = 4$ theories admit deformations to theories with only $\mathcal{N} = 2$ supersymmetry, such as turning on complex masses, FI parameters and/or superpotentials. With less supersymmetry, one typically has less control of various dualities, including 3d mirror symmetry. However it remains an interesting yet challenging arena to explore. In the following, we will discuss two examples of 3d $\mathcal{N} = 2$ mirror symmetry, at the level of superconformal indices, generalized to the case with intersecting space $S^2_{(\text{L})} \times S^1 \cup S^2_{(\text{R})} \times S^1$. The first example is the generalization of the basic duality between the $U(1)_{k = \frac{1}{2}}$ theory coupled to a fundamental chiral and the theory of a free chiral multiplet. The second example is to reduce the S-duality, \emph{i.e.}, the fiber-base duality, to the 3d $\mathcal{N} = 2$ mirror symmetry between a class of simple 3d theories on $S^2_{(\text{L})} \times S^1 \cup S^2_{(\text{R})} \times S^1$ through the Higgsing procedure. The identical indices of these theories on $S^3_\text{L} \cup S^3_\text{R}$ are taken as an evidence of mirror duality.

\subsubsection{\texorpdfstring{$U(1)_{\frac{1}{2}}$}{} theory}

Let us first recall the duality between a $U(1)_{k = \frac{1}{2}}$ theory coupled to one fundamental chiral multiplet (fund) and the theory of a free chiral multiplet. At the level of indices, one has
\begin{align}
	I_{U(1)_{\frac{1}{2}}+\text{fund}} = \oint \frac{dz}{2\pi i z} (-z)^{ - \frac{1}{2}B} (-w)^B ( - z^{-1} t\tau q^{ - \frac{1}{2}}) \frac{(z (t \tau)^{-1}q^{1 - \frac{B}{2}};q)}{(z^{-1}t\tau q^{-\frac{B}{2}};q)} = & \ \frac{((t\tau q^{-\frac{1}{2}})^{\frac{1}{2}}w^{-1}q;q)}{((t\tau q^{-\frac{1}{2}})^{-\frac{1}{2}}w;q)} \nonumber \\
	= & \ I_\text{chiral} \ .
\end{align}
Note that this basic duality between the two theories can be used to induce the order-3 $ST \in SL(2,\mathbb{Z})$ action on the $U(1)_k$ theory coupled to a fundamental chiral \cite{Krattenthaler:2011da,Cheng:2020zbh}. To see this, we introduce a background Chern-Simons term $z^{-\tilde B}$ with unit Cherns-Simons level on the right, and the above is refined to
\begin{align}
	I_{U(1)_{\frac{1}{2}} + \text{fund}}(\tilde B)
	= (t\tau)^{- \tilde B} \frac{( (t\tau q^{-\frac{1}{2}})^{\frac{1}{2}}w^{-1} q^{1 - \frac{\tilde B}{2}} ;q)}{( (t\tau q^{-\frac{1}{2}})^{- \frac{1}{2}} w q^{- \frac{\tilde B}{2}} ;q)} \ ,
\end{align}
meaning that any chiral multiplet contribution in the index of an interacting theory can be effectively replaced by an $U(1)_{\frac{1}{2}}$ theory coupled to one fundamental chiral. In particular, the index of a $U(1)_k$ theory coupled to a fundamental chiral is
\begin{align}
	I_{U(1)_{k} + \text{fund}} = \sum_{\tilde B} \oint \frac{d\tilde z}{2\pi i \tilde z} (- \tilde z)^{-k\tilde B}(-w)^{\tilde B}(-\tilde z^{-1} t\tau q^{-\frac{1}{2}})^{\frac{\tilde B}{2}} \frac{( (t\tau q^{-\frac{1}{2}})^{\frac{1}{2}}w^{-1} q^{1 - \frac{\tilde B}{2}} ;q)}{( (t\tau q^{-\frac{1}{2}})^{- \frac{1}{2}} w q^{- \frac{\tilde B}{2}} ;q)} \ ,
\end{align}
where the last factor can be replaced by $I_{U(1)_{\frac{1}{2}} + \text{fund}}(\tilde B)$. The sum over $\tilde B$ and the integral over $\tilde z$ can be easily performed, giving
\begin{align}
	I_{U(1)_{k} + \text{fund}}(w) = I_{U(1)_{\tilde k} + \text{fund}}(\tilde w) \ ,
\end{align}
where
\begin{align}
	\tilde k = \frac{2k-3}{4k+2}, \qquad \tilde w = \left[ (t\tau)^3 q^{-\frac{1}{2}} \right]^{\frac{2k-1}{4k+2}} w^{- \frac{2}{2k+1}}e^{\frac{2k-5}{2k+1} i \pi} \ .
\end{align}
Defining the map $ST: (k, w) \to (\tilde k, \tilde w)$, although the expression for $\tilde w$ looks complicated, it is straightforward to check that indeed $(ST)^3 = 1$.

Let us now consider the intersecting $U(1)_{k = \frac{1}{2}}$ theory coupled to one fundamental chiral on $S^2_{(\text{L})} \times S^1 \cup S^2_{(\text{R})} \times S^1$ with additional bifundamental 1d chiral multiplets. The index can be computed by summing over four types of poles,
\begin{center}
	\begin{tabular}{c c c}
		\text{old} & $z_{(\alpha)} \to (t\tau q^{\frac{1}{2}(\mathfrak{m} + \overline{\mathfrak{m}})})_{(\alpha)}$ & $B_{(\alpha)} \to (\mathfrak{m} - \overline{\mathfrak{m}})_{(\alpha)}$ \\
		\hline
		\text{N} & $z_{(\text{L})} \to (t\tau q^{\frac{1}{2}(-1 + \overline{\mathfrak{m}})})_{(\text{L})}$ &
		$B_{(\text{L})} \to -1 - \overline{\mathfrak{m}}_{(\text{L})}$\\
		& $z_{(\text{R})} \to (t\tau q^{\frac{1}{2}(\overline{\mathfrak{m}})})_{(\text{R})}$ &
		$B_{(\text{L})} \to - \overline{\mathfrak{m}}_{(\text{R})}$\\
		\hline
		\text{S} & $z_{(\text{L})} \to (t\tau q^{\frac{1}{2}( \mathfrak{m} -1)})_{(\text{L})}$ &
		$B_{(\text{L})} \to \mathfrak{m}_{(\text{L})} - (-1)$\\
		& $z_{(\text{R})} \to (t\tau q^{\frac{1}{2}(\mathfrak{m})})_{(\text{R})}$ &
		$B_{(\text{R})} \to (\mathfrak{m})_{(\text{R})}$\\
		\hline
		\text{NS} & $z_{(\text{L})} \to (t\tau q^{-1})_{(\text{L})}$ &
		$B_{(\text{L})} \to 0$\\
		& $z_{(\text{R})} \to (t\tau)_{(\text{R})}$ &
		$B_{(\text{R})} \to 0$
	\end{tabular}
\end{center}
The full residue can be organized into a factorized form,
\begin{align}
	= & \ \left[
	  \frac{(-w)^{-1}(t\tau q^{-1/2})^{\frac{1}{2}}_{(\text{L})}}{
	    q_{(\text{L})}^{-\frac{1}{2}}{q_{(\text{R})}^{-\frac{1}{2}}}
	    - q_{(\text{L})}^{\frac{1}{2}}{q_{(\text{R})}^{\frac{1}{2}}}
	  }
	  + \sum_{\mathfrak{m}_{(\alpha)} = 0}^{+\infty} Z_\text{vortex}(\mathfrak{m}|q)_{(\text{L})}Z_\text{vortex}(\mathfrak{m}|q)_{(\text{R})} Z_\text{intersection}(\mathfrak{m}_{(\text{L})}, \mathfrak{m}_{(\text{R})}) 
	\right]\\
	& \times \left[
	  \frac{(-w)(t\tau q^{-1/2})^{ - \frac{1}{2}}_{(\text{L})}}{
	    q_{(\text{L})}^{\frac{1}{2}}{q_{(\text{R})}^{\frac{1}{2}}}
	    - q_{(\text{L})}^{-\frac{1}{2}}{q_{(\text{R})}^{-\frac{1}{2}}}
	  }
	  + \sum_{\overline{\mathfrak{m}}_{(\alpha)} = 0}^{+\infty} Z_\text{vortex}(\overline{\mathfrak{m}}|q^{-1})_{(\text{L})}Z_\text{vortex}(\overline{\mathfrak{m}}|q^{-1})_{(\text{R})} Z_\text{intersection}(\overline{\mathfrak{m}}_{(\text{L})}, \overline{\mathfrak{m}}_{(\text{R})}) 
	\right]\ , \nonumber
\end{align}
where
\begin{align}
	Z_\text{vortex}(\mathfrak{m}|q)\equiv \frac{(t\tau q^{-\frac{1}{2}})^{-\frac{\mathfrak{m}}{2}} w^\mathfrak{m}}{(q;q)_\mathfrak{m}} \ .
\end{align}
It is straightforward to check that the above index $I_{U(1)_{\frac{1}{2}}+\text{fund}}$ equals the index of the theory of free chirals on $S^2_{(\text{L})} \times \cup S^1 \cup S^2_{(\text{R})} \times S^1$,
\begin{align}
	I^{1 \cup 2}_{U(1)_{\frac{1}{2}}+\text{fund}} = I^1_\text{chiral}I^2_\text{chiral}I_\text{1d} \ ,
\end{align}
where the indices on the right are given by
\begin{align}
	I^{\alpha}_\text{chiral} = \frac{((t\tau q^{-\frac{1}{2}})^{\frac{1}{2}}w^{-1}q;q)_{(\alpha)}}{((t\tau q^{-\frac{1}{2}})^{-\frac{1}{2}}w;q)_{(\alpha)}} \ ,
	\qquad
	I_\text{1d} = \frac{(w^{-\frac{1}{2}}\mu^{-1}_\text{f} - w^{\frac{1}{2}}\mu_\text{f})(w^{\frac{1}{2}}\mu_\text{f} - w^{- \frac{1}{2}}\mu^{-1}_\text{f})}{(q_{(\text{L})}q_{(\text{R})} - q^{-1}_{(\text{L})}q^{-1}_{(\text{R})})(q^{-1}_{(\text{L})}q^{-1}_{(\text{R})} - q_{(\text{L})}q_{(\text{R})})}\ ,
\end{align}
with the fugacity $\mu_\text{f} \equiv \prod_{\alpha = 1}^{2}(t \tau q^{-1/2})_{(\alpha)}^{ - 1/4}$. Under this duality, the 1d bifundamental chiral multiplets in the SQED on the left are dual to a pair of free 1d fermi and chiral multiplets on the right. It would be interesting to generalize to dualities for intersecting $U(1)_k$ theories using the one with both $k = 1/2$.

\subsubsection{3d mirror symmetry from fiber-base duality}

Next we turn to the 3d consequence of the fiber-base duality between 5d $\mathcal{N} = 1$ SQED and some free theory. They can be realized by simple $(p,q)$-brane web, and their full partition function on $\mathbb{R}^4 \times S^1$ can be effectively computed using the refined topological string. Here we follow the formalism and convention proposed in . Concretely, the fiber-base duality states that the SQED full partition function
\begin{align}
	Z_\text{SQED}^{\mathbb{R}^4 \times S^1} = \prod_{i = 1}^2 \frac{1}{(Q_i \mathfrak{p q}^{1/2};\mathfrak{p}, \mathfrak{q})}\sum_Y (\mathfrak{pq}^{-1/2}Q_0)^|Y|
	\frac{
	N_{\empty Y}(Q_1 (\mathfrak{pq})^{1/2};\mathfrak{p},\mathfrak{q})
	N_{\empty Y}(Q_2 (\mathfrak{pq})^{1/2};\mathfrak{p},\mathfrak{q})
	}{N_{\lambda\lambda(1; \mathfrak{p}, \mathfrak{q})}} \ .
\end{align}
equals the partition function \cite{Awata:2008ed}
\begin{align}
	Z^{\mathbb{R}^4\times S^1}_\text{free} = \left[\prod_{i = 1}^2 \frac{1}{(Q_i (\mathfrak{p q})^{1/2};\mathfrak{p}, \mathfrak{q})}\right]
	\frac{
	  (Q_0Q_1; \mathfrak{p}, \mathfrak{q})
	  (Q_0Q_2\mathfrak{pq}; \mathfrak{p}, \mathfrak{q})
	}{
	  (Q_0 (\mathfrak{pq})^{1/2};\mathfrak{p},\mathfrak{q})
	  (Q_0Q_1Q_2 (\mathfrak{pq})^{1/2};\mathfrak{p},\mathfrak{q})
	} \ .
\end{align}
Namely,
\begin{align}
	Z_\text{SQED}^{\mathbb{R}^4\times S^1} = Z^{\mathbb{R}^4\times S^1}_\text{free} \ .
\end{align}
Here $Q_{1,2}$ encode the flavor and gauge fugacities and $Q_0$ encodes the gauge coupling constant
\begin{align}
	Q_1 = z^{-1}\tilde M, \qquad Q_2 = z M^{-1}, \qquad Q_0 = - e^{- \frac{8\pi^2\beta}{g^2_\text{YM}}} M^{1/2}\tilde M^{-1/2} \ .
\end{align}
The full 5d index of the SQED is given by a contour integral of $|Z_\text{SQED}^{\mathbb{R}^4 \times S^1}|^2$ where the modular square inverts all fugacities including the $\Omega$-deformation parameters $\mathfrak{p,q}$. The Higgsing procedure on both sides picks up residues at the poles
\begin{align}
	z \to M \mathfrak{p}^{- n^\text{L} - 1/2}\mathfrak{q}^{- n^\text{R} - 1/2}\ .
\end{align}
As we have discussed in the previous section, the residue on one side of the equality organizes into the index of a $U(n^\text{L}) \times U(n^\text{R})$ gauge theory on the intersecting space $S^2_\text{L} \times S^1 \cup S^2_\text{R} \times S^1$ coupled to one pair of fundamental and anti-fundamental chiral multiplets on each $S^2 \times S^1$, and additional pair of 1d chiral multiplets in the bifundamental representation under the gauge group $U(n^\text{L}) \times U(n^\text{R})$. On the other side, $|Z^{\mathbb{R}^4 \times S^1}_\text{free}|^2$ reduces to the index of a collection of free chiral multiplets together with 1d contributions from the Fermi multiplets localized at the north and south circles,
\begin{align}
	& \ \prod_{k = 1}^{n^\text{L}}\prod_{\ell = 1}^{n^\text{R}}
	\left(1 - w (M \tilde M^{-1})^{1/2}\mathfrak{p}^{\frac{1}{2} - k}\mathfrak{q}^{\frac{1}{2} - \ell}\right)
	\prod_{\mu = 0}^{n^\text{L} - 1}\prod_{\nu = 0}^{n^\text{R} - 1}\left(1 - w (M \tilde M^{-1})^{ - 1/2}\mathfrak{p}^{\frac{1}{2} + \mu}\mathfrak{q}^{\frac{1}{2} + \nu}\right) \nonumber \\
	\times & \ \frac{1}{\prod _{\mu=0}^{n^\text{L} - 1} \prod _{\nu=0}^{n^\text{R} - 1} \left(1-\mathfrak{p}^{- \mu - 1} \mathfrak{q}^{- \nu - 1}\right) \prod _{\mu=0}^{n^\text{L}-1} \prod _{\nu=0}^{n^\text{R}-1} (1-M^{-1}\tilde M \mathfrak{p}^{\mu+1} \mathfrak{q}^{\nu+1})}\nonumber\\
	\times & \ \prod_{\mu = 0}^{n^\text{L} - 1}\frac{
	  (w (M \tilde M^{-1})^{\frac{1}{2}} (\mathfrak{pq})^{\frac{1}{2}} \mathfrak{p}^{- \mu - 1} ; \mathfrak{q})}{
	  (w (M \tilde M^{-1})^{-\frac{1}{2}} (\mathfrak{pq})^{\frac{1}{2}}\mathfrak{p}^\mu; \mathfrak{q})
  }
  \prod_{\nu = 0}^{n^\text{R} - 1}\frac{
	  (w (M \tilde M^{-1})^{\frac{1}{2}} (\mathfrak{pq})^{\frac{1}{2}} \mathfrak{q}^{- \nu - 1} ; \mathfrak{p})}{
	  (w (M \tilde M^{-1})^{-\frac{1}{2}} (\mathfrak{pq})^{\frac{1}{2}}\mathfrak{q}^\nu; \mathfrak{p})
  } \\
  \times & \ \frac{
	  \prod _{\mu=0}^{n^\text{L}-1} (M^{-1}\tilde M \mathfrak{q} \mathfrak{p}^{\mu+1};\mathfrak{q})
	 }{
	   \prod _{\mu=0}^{n^\text{L} - 1} (\mathfrak{p}^{- \mu - 1};\mathfrak{q})
	 }
	 \frac{
	   \prod _{\nu=0}^{n^\text{R}-1} (M ^{-1}\tilde M \mathfrak{p} \mathfrak{q}^{\nu+1};\mathfrak{p})
	 }{
	 	 \prod _{\nu=0}^{n^\text{R} - 1} (\mathfrak{q}^{-\nu - 1};\mathfrak{p})
 	 }\nonumber\\
 	 \times & \ (\text{fugacities} \to \text{fugacities}^{-1}) \ .\nonumber
\end{align}
Here $w \equiv -e^{- \frac{8\pi^2 \beta}{g_\text{YM}^2}}$, and the 5d flavor fugacities $M$, $\tilde M$ can be further rewritten in terms of the 3d fugacities using (\ref{5d-3d-fugacities}).

Let us go through a few simple instances in the above dualities. We start with the simplest case where $n^\text{L} = 1, n^\text{R} = 0$. The residue of the 5d SQED index gives the 3d index of the SQED theory in the factorized form (with a decoupled factor $[(v^{-1}q;q)/(v;q)]_\text{L}$ from the $U(n^\text{L})$ adjoint chiral to be omitted),
\begin{align}
	I_\text{SQED} = Z_\text{1-loop}\Big|Z_\text{vortex}\Big|^2 \ ,
\end{align}
where
\begin{align}
	Z_\text{1-loop} & \ = \left[\frac{(v^{-1}q;q)}{(v;q)}
		  \frac{((t\tilde t \tau^2)^{-1}q;q)}{((t\tilde t \tau^2);q)}\right]_\text{L}\\
	Z_\text{vortex} & \ = \sum_{\mathfrak{m} = 0}^{+\infty}(w (t\tilde t\tau^2 q^{-1})^{- \frac{1}{2}})_\text{L}^\mathfrak{m} \prod_{k = 0}^{\mathfrak{m} - 1}\left[\frac{(t \tilde t \tau^2; q)_\mathfrak{m}}{(q;q)_\mathfrak{m}}\right]_\text{L} \ .
\end{align}
On the other hand, the residue from the $Z^\text{5d}_\text{free}$ gives the index of the standard XYZ model,
\begin{align}
  I^\text{L}_\text{XYZ} = & \ 
  \left[
  \frac{((t\tilde t \tau^2)^{-1}q;q)}{((t\tilde t \tau^2);q)}\right]_\text{L}
  \left[\frac{( (t \tilde t \tau^2)^{\frac{1}{2}}w q^{\frac{1}{2}};q)}{( (t \tilde t \tau^2)^{ - \frac{1}{2}}w q^{\frac{1}{2}};q)}\right]_\text{L}
  \left[\frac{( (t \tilde t \tau^2)^{\frac{1}{2}}w^{-1} q^{\frac{1}{2}};q)}{( (t \tilde t \tau^2)^{ - \frac{1}{2}}w^{-1} q^{\frac{1}{2}};q)} \right]_\text{L}\\
  = & \ \left[
  \frac{((t\tilde t \tau^2)^{-1}q;q)}{((t\tilde t \tau^2);q)}\right]_\text{L}
  \left[\frac{( (t \tilde t \tau^2)^{\frac{1}{2}}w^{-1} q^{\frac{1}{2}};q)}{( (t \tilde t \tau^2)^{ - \frac{1}{2}}w q^{\frac{1}{2}};q)}\right]_\text{L}
  \left[\frac{( (t \tilde t \tau^2)^{\frac{1}{2}}w q^{\frac{1}{2}};q)}{( (t \tilde t \tau^2)^{ - \frac{1}{2}}w^{-1} q^{\frac{1}{2}};q)} \right]_\text{L} 
  =  I^\text{L}_Z I^\text{L}_X I^\text{L}_Y\ .
\end{align} 
It is well-known that the indices of the two theories are equal, as guaranteed by the 3d mirror symmetry. More explicitly, it is due to the identity
\begin{align}
	\sum_{\mathfrak{m} = 0}^{+\infty} \frac{(a;q)_\mathfrak{m}}{(q;q)_\mathfrak{m}} z^\mathfrak{m} = \frac{(az;q)}{(z;q)} \ ,
\end{align}
with $z = ((t \tilde t \tau)^{-1/2} w^\pm q^{1/2})_\text{L}$, $a = (t \tilde t \tau^2)^\pm_\text{L}$ and $q = q_\text{L}^\pm$. One may encode the charges of the X, Y, and Z chiral multiplets in
\begin{align}
 	m_X = (t \tilde t \tau^2 q^{-1})^{-1/2} w, \qquad
 	m_Y = (t \tilde t \tau^2 q^{-1})^{-1/2} w^{-1}, \qquad 
 	m_Z = (t \tilde t \tau^2) \ ,
\end{align} 
which satisfy the superpotential constraint $m_X m_Y m_Z = q$.

Next we consider the case with $n^\text{L} = n^\text{R} = 1$. As claimed in the previous discussions, the left hand side reduces to the intersecting index of the $U(1) \times U(1)$ SQCDA theory on $S^2_\text{L} \times S^1 \cup S^2_\text{R} \times S^1$ coupled to some 1d chiral multiplets. Indeed, the index can be computed by a JK-residue prescription applied to the contour integral
\begin{align}
	I^{\text{L}\cup \text{R}}_{\text{SQCDA}} = & \ \sum_{B_\text{L}, B_\text{R} \in \mathbb{Z}} \oint \prod_{\alpha = \text{L,R}}\frac{dz_{(\alpha)}}{2\pi i z_{(\alpha)}} (-w_{(\alpha)})^{B_{(\alpha)}} \frac{1}{(v_{(\alpha)};q)(v_{(\alpha)};q_{(\alpha)}^{-1})} \nonumber\\
  & \ \times \prod_{\alpha = \text{L,R}}\left(z^{-B}(t \tilde t^{-1})^{\frac{B}{2}}\right)_{(\alpha)} \nonumber
    \frac{(z (t\tau)^{-1}q^{1 - \frac{B}{2}};q)_{(\alpha)}}{(z^{-1} (t\tau)q^{- \frac{B}{2}};q)_{(\alpha)}}\nonumber
    \frac{(z^{-1} (\tilde t\tau)^{-1}q^{1 + \frac{B}{2}};q)_{(\alpha)}}{(z (\tilde t\tau)q^{+ \frac{B}{2}};q)_{(\alpha)}}\\
  & \ \times \prod_{\pm} \frac{1}{\sqrt{(zq^{\frac{B}{2}})_\text{L}(z^{-1}q^{- \frac{B}{2}})_\text{R} q_\text{L}^{\pm 1/2}q_\text{R}^{\pm 1/2} } - \sqrt{(z^{-1}q^{ - \frac{B}{2}})_\text{L}(zq^{\frac{B}{2}})_\text{R} q_\text{L}^{\mp 1/2}q_\text{R}^{ \mp 1/2} }  }\\
  & \ \times \prod_{\pm} \frac{1}{\sqrt{(zq^{ - \frac{B}{2}})_\text{L}(z^{-1}q^{\frac{B}{2}})_\text{R} q_\text{L}^{\pm 1/2}q_\text{R}^{\pm 1/2} } - \sqrt{(z^{-1}q^{\frac{B}{2}})_\text{L}(zq^{ - \frac{B}{2}})_\text{R} q_\text{L}^{\mp 1/2}q_\text{R}^{ \mp 1/2} }  } \ ,\nonumber
\end{align}
where the fugacities satisfy the relations
\begin{align}\label{fugacity-relations}
	& w_\text{L} = w_\text{R}, \quad
	v_\text{L} = q_\text{R}^{-1}, \quad
	v_\text{R} = q_\text{L}^{-1}, \quad
	(t\tau q^{-1/2})_\text{L} = (t\tau q^{-1/2})_\text{R} \ ,\quad
	(\tilde t\tau q^{-1/2})_\text{L} = (\tilde t\tau q^{-1/2})_\text{R}\ .
	\nonumber
\end{align}

The JK-residue prescription picks out four set of poles. The first set are of type old, given by
\begin{align}
  (z q^{ \frac{B}{2}})_{(\alpha)} = (t \tau q^{\mathfrak{m}})_{(\alpha)}, \qquad
  (z q^{ - \frac{B}{2}})_{(\alpha)} = (t \tau q^{\overline{\mathfrak{m}}})_{(\alpha)} \ .
\end{align}
Poles of type N are given by
\begin{align}
  &(zq^{\frac{B}{2}})_{\text{R}} =  (t\tau)_{\text{R}}, 
  & & (zq^{- \frac{B}{2}})_{\text{R}} = (t\tau q^{\overline{\mathfrak{m}}})_{\text{R}} \ ,
  & &
  \overline{\mathfrak{m}}_\text{R} \ge 0 \ ,
  \\
  &(z q^{\frac{B}{2}})_{\text{L}} = (t\tau q^{-1})_{\text{L}} \ ,
  & &
  (zq^{- \frac{B}{2}})_{\text{L}} = (t\tau q^{\overline{\mathfrak{m}}})_{\text{L}} \ , & &
  \overline{\mathfrak{m}}_\text{L} \ge 0 \ .
\end{align}
These poles obviously satisfy, thanks to the above fugacity relations (\ref{fugacity-relations}),
\begin{align}
	(z q^{\frac{B}{2}})_\text{L} (zq^{\frac{B}{2}})^{-1}_\text{R} (q_\text{L}q_\text{R})^{1/2} - 1
	= (t\tau q^{-1})_\text{L}(t\tau)_\text{R}^{-1}(q_\text{L}q_\text{R})^{1/2} - 1
	= 0 \ ,
\end{align}
which corresponds a simple pole of the 1d contribution in the contour integral. Similarly, poles of type S are given by
\begin{align}
  & (zq^{\frac{B}{2}})_\text{R} = (t \tau q^\mathfrak{m})_\text{R}\ ,
  && (zq^{- \frac{B}{2}})_\text{R} = (t \tau)_\text{R} \ ,
  && \mathfrak{m}_\text{R} \ge 0 \ ,\\
  & (zq^{\frac{B}{2}})_\text{L} = (t \tau q^{\mathfrak{m}})_\text{L} \ ,
  && (zq^{- \frac{B}{2}})_\text{L} = (z \tau q^{-1})_\text{L}
  && \mathfrak{m}_\text{L} \ge 0 \ .
\end{align}
Finally, theres is one pole of type NS, given by
\begin{align}
  & (z q^{ \frac{B}{2}})_\text{R} = (t \tau)_\text{R},
  && (z q^{ - \frac{B}{2}})_\text{R} = (t \tau)_\text{R} \\
  & (z q^{\frac{B}{2}})_\text{L} = (t \tau q^{-1})_\text{L},
  && (z q^{ - \frac{B}{2}})_\text{L} = (t \tau q^{-1})_\text{L} \ .
\end{align}
Note that this pole is a double pole of the 1d contribution alone, however, it also leads to a first order zero in the fundamental chiral contribution on $S^2_\text{L} \times S^1$, since
\begin{align}
  \frac{(z (t \tau)^{-1}q^{1 - \frac{B}{2}};q)_\text{L}}{(z^{-1} (t \tau)q^{- \frac{B}{2}};q)_\text{L}} \xrightarrow{\text{type NS}} \frac{(1; q)}{(q;q)} \sim 0 \ .
\end{align}
They combine to produce a simple pole of the full integrand. More precisely, one uses
\begin{align}
	\operatorname{Res}_{z \to 1} \frac{(z;q)}{(q;q)} \frac{1}{(z^{1/2} - z^{-1/2})^2}
	= \operatorname{Res}_{z \to 1} \frac{(zq;q)}{(q;q)}  \frac{(1 - z)}{(z^{1/2} - z^{-1/2})^2}
	= -1 \ .
\end{align}

Collecting the residues from all four type of poles, we have the index of the intersecting $U(1) \times U(1)$ SQCDA
\begin{align}
	\label{instersecting-SQED-index}
	I^{\text{L}\cup \text{R}}_{\text{SQCDA}} = Z_\text{1-loop}
	\Bigg|
	Z_\text{semi-vortex}& Z_\text{intersection}(\mathfrak{m}_\text{L} = -1, \mathfrak{m}_\text{R} = 0)\nonumber \\
	& \ + \sum_{\mathfrak{m}_\text{R,L} = 0}^{+\infty}Z_\text{vortex}(\mathfrak{m};q)_\text{L}Z_\text{vortex}(\mathfrak{m};q)_\text{R}Z_\text{intersection}(\mathfrak{m}_\text{L}, \mathfrak{m}_\text{R})\Bigg|^2 \ ,
\end{align}
where
\begin{align}
	Z_\text{1-loop}
	= \prod_{\alpha = \text{L,R}}^{} \left[\frac{(v^{-1}q;q)}{(v;q)}
    \frac{((t\tilde t \tau^2)^{-1}q;q)}{((t\tilde t \tau^2);q)}\right]_{(\alpha)} \ ,
\end{align}
and
\begin{align}
	Z_\text{vortex}(\mathfrak{m};q) = & \ (w (t\tilde t\tau^2 q^{-1})^{- \frac{1}{2}})^\mathfrak{m} \prod_{k = 0}^{\mathfrak{m} - 1}\left[\frac{(t \tilde t \tau^2; q)_\mathfrak{m}}{(q;q)_\mathfrak{m}}\right]\\
	(Z_\text{semi-vortex}Z_\text{intersection})(\mathfrak{m}_\text{L} = -1, \mathfrak{m}_\text{R} = 0) 
	= & \ \left[\frac{ (- w)_\text{L}^{-1} (t\tilde t\tau q^{-1})_\text{L}^{1/2} }{(1 - t \tilde t \tau^2 q^{-1})_\text{L}}\right]
	\frac{1}{(q_\text{L}q_\text{R})^{-\frac{1}{2}} - (q_\text{L}q_\text{R})^{\frac{1}{2}}} \ .
\end{align}

Thanks to the S-duality, the index (\ref{instersecting-SQED-index}) can be reorganized into the following (ignoring the factor $\prod_{\alpha = \text{L,R}}[\frac{(v^{-1}q;q)}{(v;q)}]_{(\alpha)}$),
\begin{align}
	I^{\text{L}\cup \text{R}}_{\text{XYZ}}
	= \Bigg[\frac{(1 - (m_X^\text{L}m_X^\text{R})^{\frac{1}{2}})(1 - (m_Y^\text{L}m_Y^\text{R})^{\frac{1}{2}})}{(1 - (v_\text{L}v_\text{R})^{\frac{1}{2}}) (1 - (m_Z^\text{L}m_Z^\text{R}v_\text{L}v_\text{R})^{-\frac{1}{2}})}\Bigg]
	\Bigg[m \to m^{-1}, v \to v^{-1}\Bigg]
	\prod_{\alpha = \text{L,R}}(I_X I_Y I_Z)_\alpha\ .
\end{align}
Here we recognize the index of two $XYZ$ models living on the intersecting $S^2_\text{L}\times S^1 \cup S^2_\text{R} \times S^1$, with additional contributions from 1d free chiral and free Fermi multiplets on the intersection $S^1$ captured by the fraction in front. The fugacities are defined naturally for $\alpha = \text{L,R}$ by
\begin{align}
 	m_X^\alpha = (t \tilde t \tau^2 q^{-1})_\alpha^{-1/2} w, \qquad
 	m_Y^\alpha = (t \tilde t \tau^2 q^{-1})_\alpha^{-1/2} w^{-1}, \qquad 
 	m_Z^\alpha = (t \tilde t \tau^2)_\alpha \ .
\end{align}

Next we consider the non-abelian but non-intersecting case of the duality with $n^\text{L}> 1, n^\text{R} = 0$. On one side, we obtain the index $I^\text{L}_{U(n^\text{L})\text{-SQCDA}}$ of a $U(n^\text{L})$ gauge theory with a pair of fundamental/anti-fundamental, and one adjoint chiral multiplets. On the other side, one has 
\begin{align}
	\widehat{I}^\text{L}_{XYZ} = \left[I_Z\prod_{\mu = 0}^{n^\text{L} - 1} I_{X_\mu} I_{Y_\mu} \right] 
	\left[
		I_\text{adj}
	  \prod_{\mu = 2}^{n^\text{L} }\frac{1}{I_{\beta_\mu}}
	\right]
	\left[
	  \prod_{\mu = 0}^{n^\text{L} - 2}\frac{1}{I_{\gamma_\mu}}
	\right] \equiv I^\text{L}_{XYZ} \frac{1}{I^\text{L}_{\beta \gamma}}\ ,
\end{align}
where we have the free chiral indices $I_\text{matter} \equiv (m^{-1}q_\text{L};q_\text{L})/(m ;q_\text{L})$ with
\begin{align}
	& m[X_\mu] = (w (t \tilde t \tau^2)^{-\frac{1}{2}} q^{\frac{1}{2}} v^{- \mu})_\text{L} \ , 
	\quad
	m[Y_\mu] = (w^{-1} (t \tilde t \tau^2)^{-\frac{1}{2}} q^{\frac{1}{2}} v^{- \mu})_\text{L}, \quad \mu = 0, 1, \ldots, n^\text{L} - 1\\
	& m[Z] = ((t \tilde t \tau^2) v^{n^\text{L} - 1})_\text{L} \ , \qquad
	m_\text{adj} = v
	\\
	& m[\gamma_{\mu = 0, \ldots, n^\text{L} - 2}] = ((t \tilde t \tau^2)^{-1}v^{- \mu}q)_\text{L}, \qquad
	m[\beta_{\mu = 2, \ldots, n^\text{L}}] = (v^{ - \mu}q)_\text{L} \ ,
\end{align}
and we have $I_{XYZ}$ to collectively denotes the contributions from $X_\mu, Y_\mu$ and $Z$. Here again we notice the fugacity relations $m_{X_\mu}m_{Y_{n^\text{L} - 1 - \mu}}m_Z = q_\text{L}$, compatible with an XYZ-type superpotential $\sum_{\mu = 0}^{n^\text{L} - 1} X_\mu Y_{(n^\text{L} - 1 - \mu)} Z$. Rearranging the factors of $I_{\beta_\mu}$ and $I_{\gamma_\mu}$ to the other side, we have
\begin{align}\label{non-abelian-duality-1}
	I_{\beta \gamma} \frac{I^\text{L}_{U(n^\text{L})\text{-SQCDA}}}{I_\text{adj}} = I^\text{L}_{XYZ} \ ,
\end{align}
where the free fields $\{\beta_\mu \}_{\mu = 2}^{n^\text{L}}$ and $\{\gamma_\mu\}_{\mu = 0}^{n^\text{L} - 2}$ have fugacities satisfying
\begin{align}
	m[\gamma_\mu] \ m[q] \ m[\tilde q] \ m[\Phi^\mu] = q, \qquad
	m[\beta_\mu] \ m[\Phi^{\mu}] = q
\end{align}
coming from the superpotential constraint
\begin{align}
	\sum_{\mu = 2}^{n^\text{L}} \operatorname{tr} \beta_\mu \Phi^\mu_\text{adj} + \sum_{\mu = 0}^{n^\text{L} - 2} \gamma_\mu \tilde Q\Phi^\mu_\text{adj}Q \ ,
\end{align}
where we denotes the fundamental and anti-fundamental chiral multiplets by $Q, \tilde Q$.

The equality (\ref{non-abelian-duality-1}) can be further refined by including a background Chern-Simons term at level 1 coupled to a $U(1)$ gauge field that weakly gauge the topological $U(1)$ symmetry in the $U(n^\text{L})$ SQCDA. As a result, the integrand on the left contains a term $(\prod_{a =  1}^{n^\text{L}}z_a)^{B_w}$. At this point, one can gauge the $U(1)$ topological symmetry on both side, namely, integrate over $w$ and sum over $B_w$, which simultaneous force $\sum_a B_a = 0$ and $\prod_{a} z_a = 1$, reducing the $U(n^\text{L})$ gauge symmetry to $SU(n^\text{L})$. The denominator $I_\text{adj}$ also reduces the contribution from the $n^\text{L}$ Cartan components of the $U(n^\text{L})$ adjoint chiral to $n^\text{L} - 1$ components of the $SU(n^\text{L})$ adjoint chiral multiplet. In the end, the equality of indices now reads
\begin{align}
	I_{\beta \gamma + SU(n^\text{L}) \text{-SQCDA}} = I_{U(1) + XY}I_Z \ ,
\end{align}

Finally we are ready to conclude the general case with $(n^\text{L} > 0, n^\text{R} > 0)$. As mentioned above, on one side we have the index $I^{\text{L} \cup \text{R}}_\text{SQCDA}$ of the intersecting unitary gauge theories coupled to 3d fundamental/anti-fundamental/adjoint chiral multiplets, and additional 1d chiral multiplets in the bifundamental representation of $U(n^\text{L}) \times U(n^\text{R})$. This is equal to
\begin{align}
	I^{\text{L} \cup \text{R}}_\text{SQCDA} = \frac{I^\text{L}_{XYZ}}{I_{\beta \gamma}^\text{L}}\frac{I^\text{R}_{XYZ}}{I_{\beta \gamma}^\text{R}} I^\text{1d} \ ,
\end{align}
where we have the contributions from the one dimensional matters given by
\begin{align}
	\prod_{\mu = 0}^{n^\text{L} - 1}\prod_{\nu = 0}^{n^\text{R} - 1}\frac{
	 \left(1 - (q^{-1}_\text{L}q^{-1}_\text{R}m[X^\text{L}_\mu]m[\beta_\mu^\text{L}]m[X^\text{R}_\nu]m[\beta_\nu^\text{R}])^{\frac{1}{2}}\right)
	\left(X \leftrightarrow Y\right)}{
	  (1 - m[\beta^\text{L}_\mu]m[\beta^\text{R}_\nu])
	  \left(1 - (m[\beta_\mu^\text{L}]m[\gamma_\mu^\text{L}] m[\beta_\nu^\text{R}]m[\gamma_\nu^\text{R}]q_\text{L}^{-1}q_\text{R}^{-1})^{\frac{1}{2}}\right)
	} \times (\text{fug} \to \text{fug}^{-1}) \ .
\end{align}

\subsubsection{Reduction to 2d}

Note that the two $S^2_\text{L,R} \times S^1$ shares the same circle $S^1$ whose length is controlled by the parameter $\beta$ entering into fugacities $q_\text{L,R}$. Therefore, sending $\beta \to 1$ effectively shrinks the $S^1$ and in the end one expects to to find the $S^2$-partition function, or in more general cases, $S^2_\text{L} \cup S^2_\text{R}$-partition function from the index computed in the previous discussions. The 3d mirror symmetry above then reduces to a 2d duality at the level of $S^2$ partition functions.

The reduction of indices is due to the limit
\begin{align}
	\lim_{q \to 1} \frac{(q^{a};q)}{(q^b;q)}(1-q)^{a-b} = \frac{\Gamma(b)}{\Gamma(a)} \ .
\end{align}
This limit was exploited in \cite{Benini:2012ui,Doroud:2012xw} to reduce the 3d index, the matter contributions therein in particular, to $S^2$-partition function by sending the radius $\beta$ of the temporal $S^1$ to zero while rescaling the fugacities properly and inserting the factors of $(1-q)$ by hand. In the case of $n_\text{f} = n_\text{af}$ and also for the adjoint chiral contributions, such factor are harmless. Defining $q = e^{2\pi i \beta}$, $z = q^{\hat z} = e^{2\pi i \beta \hat z}$, and similarly for other fugacities, one has the needed factors for the fundamental and anti-fundamental chirals in a $U(n)$ gauge theory given by
\begin{align}
	\prod_{a = }^{n}\prod_{i=1}^{n_\text{f}}(1 - q)^{2\hat z_a - 2(\hat t_i + \hat \tau) + 1 } (1 - q)^{-2\hat z_a - 2 (\hat {\tilde t}_i + \hat \tau) + 1}
	= \prod_{a = }^{n}\prod_{i=1}^{n_\text{f}}(1-q)^{-2(\hat t_i + \hat {\tilde t}_i + 2 \hat \tau - 1)}\ ,
\end{align}
and for the adjoint
\begin{align}
	\prod_{a \ne b} (1 - q)^{2\hat z_a - 2\hat z_b - 2\hat v + 1}
	= \prod_{a \ne b} (1 - q)^{- 2\hat v + 1} \ .
\end{align}
These factors are simple constant factors that can be taken out of the integral over $z$'s and the sum over $B$'s.

This procedure straightforwardly carries over to the intersecting case, where the chiral multiplets and the vector multiplet contributions in the index reduce to those in the $S^2_\text{L}$- and $S^2_\text{R}$-partition functions. The remaining factors to reduce are the contributions from the 1d matters. Let us parametrize, using $\epsilon_1, \epsilon_2$ which originates from the 5d $\Omega$-deformation parameters,
\begin{align}
	q_\text{L} = e^{2\pi i \beta \epsilon_2}, \qquad q_\text{R} = e^{2\pi i \beta \epsilon_1} \ ,
\end{align}
and define the hatted fugacities, for example,
\begin{align}
	z^{\text{L}}_a = e^{2 \pi i \beta \epsilon_2\hat z_a^\text{L}},
	\quad
	q_\text{L}^{\frac{1}{2}B_a^\text{L}} = e^{\pi i \beta \epsilon_2 B_a^\text{L}}\ ,
	\qquad
	z^\text{R}_a = e^{2\pi i \beta \epsilon_1 \hat z_a^\text{R}} \ ,\quad
	q_\text{R}^{\frac{1}{2}B_a^\text{R}} = e^{\pi i \beta \epsilon_2 B_a^\text{R}} \ .
\end{align}
In this parametrization, $\beta$ encodes the radius of the common $S^1$. The contribution from the 1d chiral multiplet then reads
\begin{align}
	& \ \prod_\pm \ \prod_{\pm '}\frac{1}{2 \sinh \pi i \beta (\epsilon_2 (\hat z_a^\text{L} \pm \frac{B_a^\text{L}}{2}) - \epsilon_1 (\hat z_a^\text{R} \pm \frac{B_a^\text{R}}{2}) \pm' \frac{1}{2}(\epsilon_1 + \epsilon_2))} \\
	\xrightarrow{\beta \to 0} \ &
	\prod_\pm \ \prod_{\pm '}\frac{1}{2 \pi i \beta (\epsilon_2 (\hat z_a^\text{L} \pm \frac{B_a^\text{L}}{2}) - \epsilon_1 (\hat z_a^\text{R} \pm \frac{B_a^\text{R}}{2}) \pm' \frac{1}{2}(\epsilon_1 + \epsilon_2))} \ ,
\end{align}
which is simply the contributions from the 0d chiral multiplets living at the north and south pole of $S^2_\text{L}$ and $S^2_\text{R}$ where they intersect.

With these observations, we can reduce the above 3d mirror symmetry descending from the S-duality to a duality between 2d $\mathcal{N} = (2,2)$ theories. At the level of partition functions on $S^2_\text{L} \cup S^2_\text{R}$, we have
\begin{align}
	Z_{\beta \gamma}Z^{\text{L} \cup \text{R}}_\text{SQCDA} = Z^\text{L}_{XYZ}Z^\text{R}_{XYZ} Z^\text{0d} \ ,
\end{align}
where
\begin{align}
	I_{XYZ}
	= &\ \frac{\Gamma( - \hat t - \hat {\tilde t} - 2 \hat \tau - (n^\text{L} - 1)\hat v + 1)}{\Gamma( \hat t + \hat {\tilde t} + 2 \hat \tau + (n^\text{L} - 1)\hat v)} \\
	& \ \times \prod_{\mu = 0}^{n^\text{L} - 1} \frac{\Gamma( - \hat w + \frac{1}{2}(\hat t + \hat {\tilde t} + 2\hat \tau) + \mu \hat v + \frac{1}{2})}{\Gamma( + \hat w - \frac{1}{2}(\hat t + \hat {\tilde t} + 2\hat \tau) - \mu \hat v + \frac{1}{2})} 
	\frac{\Gamma( + \hat w + \frac{1}{2}(\hat t + \hat {\tilde t} + 2\hat \tau) + \mu \hat v + \frac{1}{2})}{\Gamma( - \hat w - \frac{1}{2}(\hat t + \hat {\tilde t} + 2\hat \tau) - \mu \hat v + \frac{1}{2})} \ ,
\end{align}
and the 1d contribution
\begin{align}
	Z^\text{1d}
	= \prod_{\mu = 0}^{n^\text{L} - 1}\prod_{\nu = 0}^{n^\text{R} - 1} \frac{
	  ( - \epsilon_2 - \epsilon_1 + \epsilon_2 (\hat m[X^\text{L}_\mu] + \hat m[\beta_\mu^\text{L}]) + \epsilon_1 (\hat m[X^\text{R}_\nu] + \hat m[\beta_\nu^\text{R}]))^2 (X\leftrightarrow Y)
	}{
    (2\epsilon_2\hat m[\beta_\mu^\text{L}] + 2\epsilon_1\hat m[\beta_\nu^\text{R}])^2(\epsilon_2(\hat m[\beta^\text{L}_\mu] + m[\gamma^\text{L}_\mu])
    + \epsilon_1(\hat m[\beta^\text{R}_\nu] + m[\gamma^\text{R}_\nu]))^2
	}
\end{align}
using
\begin{align}
	(1 - e^{2\pi i \beta x})(1 - e^{-2\pi i \beta x})
	= 4 (\sin \pi \beta x)^2 \xrightarrow{\beta \to 0} 4 \pi^2 \beta^2 x^2 \ .
\end{align}

\subsection{3d/3d correspondence}

A class of 3d $\mathcal{N} = 2$ supersymmetric gauge theories $\mathcal{T}[M, G]$ can be engineered by compactifying from 6d $(0,2)$ theory of type $G$ on a three manifold $M$ \cite{Dimofte:2011py,Dimofte:2011ju,Dimofte:2011py}. This construction identifies the supersymmetric vacua of $\mathcal{T}[M, G]$ defined on $\mathbb{C} \times S^1$ with the moduli space of flat $G$-connection on $M$, while the partition function of $\mathcal{T}[M,G]$ on Lens spaces $L(k,1)_b$ with the partition function of the complex refined Chern-Simons theory on $M$ at (complex) level $(k, \sigma)$,
\begin{align}
	Z^{D^2 \times S^1}_{\mathcal{T}[M, G]} = Z^M_{G\text{-CS}_{k,\sigma}} \ .
\end{align}
In the case with $M = S^3$, the $\mathcal{T}[S^3, U(N)]$ theory is given by $U(N)$ gauge theory with an adjoint chiral, and the duality has been checked via another duality from $\mathcal{T}[S^3, U(N)]$ to a collection of free chiral multiplets \cite{Gukov:2016gkn},
\begin{align}
	\label{triality}
	Z^{D^2 \times S^1}_{\mathcal{T}[S^3,U(N)]} = Z^{D^2\times S^1}_\text{free} = Z^{S^3}_{\text{CS}_{k, \sigma}} \ .
\end{align}
Concretely, one has
\begin{align}
	Z^{D^2 \times S^1}_{\mathcal{T}[S^3, U(N)]} \equiv & \ \frac{1}{N!} \oint \prod_{a = 1}^{N}\frac{dz_a}{2\pi i z_a} \frac{\prod_{a\ne b} (z_a/z_b;q_1)}{\prod_{a, b = 1}^N(z_a/z_b q_2;q_1)} \prod_{a = 1}^N \Theta(z_a;q_1) \\
	= & \ \prod_{\ell = 1}^N \frac{1}{(q_2^\ell;q_1)} = Z_\text{free}^{D^2 \times S^1}\ ,
\end{align}
where the Theta function is defined by $\Theta(z) \equiv \sum_{n\in \mathbb{Z}}q^{\frac{n^2}{2}}z^n$, capturing the contribution from the boundary degrees of freedom. The latter agrees with the refined Chern-Simons partition function
\begin{align}
	Z^{S^3}_{U(N)_k \text{def-CS}} = \left(\frac{4\pi^2}{k}\right)^{-\frac{N}{2}} \frac{1}{N!(q_2;q_1)^N} \int_{-\infty}^{+\infty} \prod_{a = 1}^{N} d\sigma_a \prod_{a \ne b}^{N} \frac{(e^{\sigma_a - \sigma_b};q_1)}{(e^{\sigma_a - \sigma_b}q_2;q_1)}e^{- \frac{k}{4\pi i} \sum_{a = 1}^N \sigma_a^2} \ .
\end{align}
For example, at the special value $q_2 = q_1^B$ with $B \in \mathbb{N}$, the partition function of the refined Chern-Simons theory with integer refinement $B$ reads
\begin{align}
	\left(\frac{4\pi^2}{k}\right)^{-\frac{N}{2}} \frac{1}{N!(q_2;q_1)^N}\int \prod_{a = 1}^{N} d\sigma_a
	\prod_{\ell = 0}^{B - 1}\prod_{a \ne b} (e^{\frac{\sigma_b - \sigma_a}{2}} - q_1^\ell e^{\frac{\sigma_a - \sigma_b}{2}}) e^{- \frac{k}{4\pi i}\sum_{a = 1}^{N} \sigma_a^2}
\end{align}
which can be rewritten into a contour integral by a change of variable $e^{\sigma_a} = x_a$, and an identity
\begin{align}
	\frac{k^{1/2}}{2\pi}\int_0^{+\infty} dx x^n e^{ - \frac{(\log x)^2}{4\pi}}
	= \oint_{|z| = 1} \frac{dz}{z}z^n \Theta(z) \ , \qquad \forall n \in \mathbb{Z} \ .
\end{align}
The resulting contour integral picks up the constant term of the integrand, and gives
\begin{align}
	Z^{S^3}_{U(N)_k \text{ def-CS}} = \prod_{\ell = 0}^{N - 1}\frac{1}{(q_1^B;q_1)}\prod_{n = 0}^{B - 1} (1 - q_1^{B\ell}q^n)^{N - \ell} = \prod_{\ell = 1}^N \frac{1}{(q_1^{B\ell};q_1)}\ ,
\end{align}
with identification
\begin{align}
	q_1 = e^{\frac{2\pi i}{k}} \ .
\end{align}

The relation between the refined Chern-Simons and $Z_\text{free}$ is also the well-known result of the open-closed duality between the open topological string on $T^*S^3$ with A-branes warpping $S^3 \subset T^*S^3$ and the topological string theory on $\mathcal{O}(-1) \oplus \mathcal{O}(-1) \to \mathbb{CP}^1$. The latter partition function is given by
\begin{align}
	Z_\text{top} = (Q;q_1, q_2^{-1})^{-1} \ ,
\end{align}
where $Q$ encodes the Kahler parameter of the $\mathbb{CP}^1$. The geometric transition can be implemented in the partition function by taking the residue at a pole $Q \to q_2^{N}$, giving $Z_\text{free}$ as the result \cite{Gukov:2016gkn},
\begin{align}
	\operatorname{Res}Z_\text{top} = \prod_{\ell = 1}^{N} \frac{1}{(q_2^{\ell}; q_1)} \ .
\end{align}
which is expected to be dual to the $S^3$ partition function of the refined $U(N)_k$ Chern-Simons.

Based on the above discussion, it is natural to consider a more general residue of the form $Q \to q_2^{ - n^\text{L}} q_1^{n^\text{R}}$. From the perspective of the brane web that engineers the pure super-Yang-Mills, the residue corresponds to suspending two set of D3 branes between the D5 and NS five branes in figure (). The result is simply
\begin{align}
	\operatorname{Res}Z_\text{top}
	= \prod_{\ell = 1}^{n^\text{L}}\frac{1}{(q_2^\ell; q_1)}
	  \prod_{\ell' = 1}^{n^\text{R}}\frac{1}{(q_1^{\ell'}; q_2)}
	  \prod_{\ell = 1}^{n^\text{L}}\prod_{\ell' = 1}^{n^\text{R}} \frac{1}{1 - q_1^{-\ell}q_2^{-\ell'}} \ .
\end{align}

In the following we focus on the simplest case $n^\text{L} = n^\text{R} = 1$ while leaving the more general situations for future study. The above residue then takes the form of free chiral multiplets on $(D^2 \times_{q_1} S^1) \cup (D^2 \times_{q_2} S^1)$ with additional 1d chiral multiplets,
\begin{align}
	\operatorname{Res}Z_\text{top} = \frac{1}{(q_2;q_1)}\frac{1}{(q_1;q_2)}\frac{1}{1 - q^{-1}_1 q^{-1}_2} \ .
\end{align}
Next we look for the corresponding gauge theory dual of this free theory, and we propose the dual to be the intersecting SQED(A) whose partition function is given by
\begin{align}
	Z_{\text{SQCDA}^2} = \oint \frac{dz_1}{2\pi i z_1}\frac{dz_2}{2\pi i z_2} \frac{1}{(q_1;q_2)}\frac{1}{(q_2;q_1)}\frac{\Theta(z_1;q_1)\Theta(z_2;q_2)}{(1 - q_1^{-1/2}q_2^{-1/2} \frac{z_1}{z_2})(1 - q_1^{-1/2}q_2^{-1/2} \frac{z_2}{z_1})} \ .
\end{align}
Indeed, treating the integrand as a Laurent series in $z_1, z_2$ in the region $|z_1| < |z_2|$, one can extract its constant terms, and obtain a $q$-series
\begin{align}
	Z_{\text{SQED(A)}^2} = - \frac{(q_1q_2)^{-\frac{1}{2}}}{(q_2; q_1)(q_1; q_2)}\sum_{\mathfrak{m}, \mathfrak{n} = 0}^{+\infty} (q_1 q_2)^{ - \frac{1}{2}(\mathfrak{m} - \mathfrak{n}) + \frac{1}{2}(\mathfrak{m}+\mathfrak{n} + 1)^2} \ .
\end{align}
The double series turns out to be given by the special value $g_{1,1,1}(-1, -x , x)$ of the false Theta function
\begin{align}
	g_{a,b,c}(x,y,q) \equiv \left(\sum_{r, s \ge 0} + \sum_{r,s < 0}\right)(-1)^{r + s} x^r y^s q^{a \frac{r(r-1)}{2} + b rs + c \frac{s(s-1)}{2}} \ .
\end{align}
In the end, the partition function evaluates to
\begin{align}
	Z_{\text{SQCDA}^2} = \frac{1}{(q_2;q_1)(q_1;q_2)} \frac{1}{1 - (q_1q_2)^{-1}}
	= \operatorname{Res}Z_\text{top} \ ,
\end{align}
reproducing the free theory index coming from the refined geometric transition.

Finally, treating the integrand of the contour integral as series in $z_1, z_2$ in the region $|z_1| < |z_2|$, one can retrace the steps which proves the (\ref{triality}), and show that
\begin{align}
	Z_{\text{SQCDA}^2} = \operatorname{Res}Z_\text{top} = \int_{-\infty}^{+\infty} d\sigma_1d\sigma_2 \frac{e^{- \frac{k_1}{4\pi i} \sigma_1^2 - \frac{k_2}{4\pi i} \sigma_2^2}}{\prod_{\pm}\sinh \pi i (\epsilon_2 \sigma_1 - \epsilon_1 \sigma_2 \pm \frac{1}{2}(\epsilon_1 + \epsilon_2))} \ .
\end{align}
The expression on the right looks like the $S^3$-partition function of two $U(1)_{k = 1}$ theories coupled through some 1d bifundamental chiral multiplets. However, the precise physical interpretation of the expression remains unclear, which we leave to future work.

\section{\texorpdfstring{$q$}{q}-Virasoro construction}

In this section we follow the idea of \cite{Nedelin:2016gwu,Nieri:2017vrb,Nieri:2018ghd,Lodin:2017lrc} using $q$-Viraosro algebra to construct and study the algebraic properties of the $\mathcal{N} = 2$ superconformal index of intersecting gauge theories. We will begin by reviewing the construction in \cite{Nedelin:2016gwu} and then generalize to theories on $S^2_\text{L}\times S^1 \cup S^2_\text{R} \times S^1$. Then we argue the uniqueness of the algebraic construction.

The $q$-Virasoro algebra $\mathbb{V}_{q,t}$ is generated by a set of infinitely many generators $\mathbf{T}_m$ satisfying the commutation relations
\begin{align}\label{qVirasoro-commutation}
  \sum_{k \ge 0}f_k (\mathbf{T}_{m - k} \mathbf{T}_{n + k}
  - \mathbf{T}_{n - k} \mathbf{T}_{m + k}) = - \frac{(1-q)(1-t^{-1})}{1-p} (p^m - p^{-m}) \delta_{m + n, 0} \ .
\end{align}
Here $p \equiv qt^{-1}$ and $f_k$ is defined by the relation $\sum_{k = 0}^{+\infty} f_k z^k = \exp \left[\sum_{m > 0}\frac{(1 - q^m)(1 - t^{-m}) z^m}{1 + p^m}\right]$. One can also pack the $\mathbf{T}_m$ in the stress tensor $\mathbf{T}(z) \equiv \sum_m \mathbf{T}_m z^{-m}$. The algebra admits a free field realization in terms of the Heisenberg operators $\{\mathbf{a}_m; \mathbf{P,Q}\}$ with commutation relations
\begin{align}
  [\mathbf{a}_m, \mathbf{a}_n] = - \frac{1}{m} (q^{m/2} - q^{- m/2}) (t^{-m/2} - t^{m/2})(p^{m/2} + p^{-m/2})\delta_{m + n, 0} , \qquad [\mathbf{P}, \mathbf{Q}] = 2 \ ,
\end{align}
and, by defining $\beta$ via $t = q^\beta$ and $: \ldots :$ to denote the normal ordering pushing $\mathbf{a}_{<0}$ to the left,
\begin{align}
  \mathbf{T}(z) = \mathbf{Y}(p^{-1/2}z) + \mathbf{Y}(p^{1/2}z)^{-1},
  \qquad
  \mathbf{Y}(z) \equiv :\exp\left[\sum_{m \ne 0} \frac{\mathfrak{a}_m z^{-m}}{p^{m/2} + p^{-m/2}}\right] q^{\sqrt{\beta}\frac{\mathbf{P}}{2}}p^{1/2}: \ .
\end{align}
An immediate observation is that the algebra $\mathbb{V}_{q, t}$ is actually invariant under $q \leftrightarrow t^{-1}$ since the commutation relation (\ref{qVirasoro-commutation}) is.

The algebra contains special operators refereed to the screening currents given by
\begin{align}
	\mathbf{S}_+(x) & \ \equiv :\exp \left[- \sum_{m \ne 0} \frac{\mathbf{a}_m x^{-m}}{q^{m/2} - q^{-m/2}} + \sqrt{\beta}\mathbf{Q}\right] z^{\sqrt{\beta}\mathbf{P}}:\\
	\mathbf{S}_-(x) & \ \equiv :\exp \left[- \sum_{m \ne 0} \frac{\mathbf{a}_m x^{-m}}{t^{-m/2} - t^{m/2}} - \sqrt{\beta}^{-1}\mathbf{Q}\right] z^{ - \sqrt{\beta}^{-1}\mathbf{P}}:\ ,
\end{align}
which satisfy
\begin{align}
	& \ [\mathbf{T}_m, \mathbf{S}_+(x)] = \frac{1}{x}\left(\mathcal{O}(q x) - \mathcal{O}(x)\right),
	\quad
	[\mathbf{T}_m, \mathbf{S}_-(x)] = \frac{1}{x}\left(\mathcal{O}(t^{-1} x) - \mathcal{O}(x)\right) \\
	& \ \Rightarrow 
	[\mathbf{T}_m, \oint dx \mathbf{S}_\pm(x)] = 0 \ ,
\end{align}
for an appropriate contour.

In \cite{Nedelin:2016gwu,Nieri:2017vrb}, two $\mathbf{S}_+$'s from two $q$-Virasoro algebras $\mathbb{V}_{q, t^{-1}}$, $\mathbb{V}_{\tilde q, \tilde t^{-1}}$ sharing the same $\beta$ are fused into a modular double screening current,
\begin{align}
	\mathbf{S}(z, \tilde z) \equiv \exp\left[
	  - \sum_{m \ne 0} \frac{\mathbf{a}_m z^{ - m}}{q^{m/2} - q^{-m/2}}
	  - \sum_{m \ne 0} \frac{\mathbf{a}_m \tilde z^{ - m}}{\tilde q^{m/2} - \tilde q^{-m/2}} - \sqrt{\beta} \mathbf{Q}
	\right] f(z, \tilde z, \sqrt{\beta}\mathbf{P}) \ ,
\end{align}
where $f(z, \tilde z, \sqrt{\beta}\mathbf{P})$ satisfies certain periodic condition under $z \to q z$ and $\tilde z \to \tilde q \tilde z$ \footnote{Note that $z$ and $\tilde z$ are not independent, and therefore the shift in $z$ will also affect $\tilde z$ in certain way, depending on the specific modular double construction.} depending on the geometry. The (sum of) normal ordered products of modular doubles then produce $S^3$ or $S^2 \times S^1$ partition functions of $\mathcal{N} = 2$ unitary gauge theories coupled to one adjoint chiral multiplet, and fundamental/anti-fundamental chiral multiplets if further shifts in the construction are implemented.

Following the logic, we consider the modular doubles $\mathcal{S}_\pm$ by merging $\mathbf{S}_\pm$ with $\tilde{\mathbf{S}}_\pm$ from the mutually commuting $q$-Virasoro algebras $\mathbb{V}_{q, t}$ and $\mathbb{V}_{\tilde q = q^{-1}, \tilde t = t^{-1}}$
\begin{align}
	\mathcal{S}_+(z,B) \equiv & \ q^{B} \exp\left[
	  - \sum_{m \ne 0} \frac{\mathbf{a}_m (z q^{\frac{B}{2}})^{- m}}{q^{\frac{m}{2}} - q^{- \frac{m}{2}}}
    - \sum_{m \ne 0} \frac{\tilde{\mathbf{a}}_m (z^{-1}q^{\frac{B}{2}})^{- m}}{\tilde q^{\frac{m}{2}} - \tilde q^{- \frac{m}{2}}}
    + \sqrt{\beta} \mathbf{Q} 
	\right]q^{B \sqrt{\beta}\mathbf{P}}\\
  \mathcal{S}_-(z,B) \equiv &\ t^{ - B}\exp\left[
  - \sum_{m \ne 0} \frac{\mathbf{a}_m (z t^{ - \frac{B}{2}})^{- m}}{t^{ - \frac{m}{2}} - t^{ \frac{m}{2}}}
  - \sum_{m \ne 0} \frac{\tilde{\mathbf{a}}_m (z^{-1}t^{ - \frac{B}{2}})^{- m}}{\tilde t^{ - \frac{m}{2}} - \tilde t^{ \frac{m}{2}}}
  - \frac{1}{\sqrt{\beta}} \mathbf{Q} 
\right] t^{ - B \sqrt{\beta}\mathbf{P}} \ ,
\end{align}
Note that the last factor in $\mathcal{S}_+$,
\begin{align}
	q^{B\sqrt{\beta}\mathbf{P}} = (zq^{\frac{B}{2}})^{\sqrt{\beta}\mathbf{P}} (z^{-1}q^{\frac{B}{2}})^{\sqrt{\beta}\mathbf{P}} \ ,
\end{align}
is the product of the $\mathbf{P}$ factor in $\mathbf{S}_+(z q^{\frac{B}{2}})$ and $\tilde{\mathbf{S}}_+(z^{-1} q^{\frac{B}{2}})$ which merge into $\mathcal{S}_+$. In other words, an $\mathcal{S}$ is essentially the product of $\mathbf{S}$ and $\tilde{\mathbf{S}}$, except that they now share the $\mathbf{Q}$ dependence. Product of such screening currents gives
\begin{align}
	& \ \prod_{a = 1}^{n^\text{L}} \mathcal{S}_+(z_a,B_a)_\text{L}\prod_{a = 1}^{n^\text{R}} \mathcal{S}_-(z_a,B_a)_\text{R} \\
	= & \ :\prod_{a = 1}^{n^\text{L}}  \mathcal{S}_+(z_a,B_a)_\text{L}\prod_{a = 1}^{n^\text{R}} \mathcal{S}_-(z_a,B_a)_\text{R}: \left[\prod_{i = \text{L,R}} Z^{(i)}_\text{VM}(z)_{(i)}Z^{(i)}_\text{adj}(z)_{(i)} \right] Z_\text{1d chiral}(z^\text{L}, B^\text{R}, z^\text{R}, B^\text{R}) \ , \nonumber
\end{align}
where (temporarily leaving the label $(i)$ implicit)
\begin{align}
	Z_\text{VM}(z) = & \ \prod_{a < b} q^{ - \frac{B_a -  B_b}{2}}(1 -  z_a z_b^{-1}q^{\frac{ B_a -  B_b}{2}})(1 -  z_b  z_a^{-1}q^{\frac{ B_a -  B_b}{2}}) \\
	Z_\text{adj}(z) = & \ \frac{(t^{-1}q;q)^n}{(t;q)^n} \prod_{a < b}(q^{-1/2}t)^{- (B_a - B_b)} \frac{(z_a^{-1}z_b t^{-1}q^{1 + \frac{B_a - B_b}{2}};q)(z_b^{-1}z_a t^{-1}q^{1 + \frac{B_a - B_j}{2}};q)}{(z_a z_b^{-1}t q^{\frac{B_a - B_b}{2}};q)(z_b z_a^{-1}t q^{\frac{B_a - B_b}{2}};q)} \ ,
\end{align}
Here we have reorganized the contribution from the adjoint chiral multiplet following \cite{Hwang:2015wna}. The 1d bifundamental chiral naturally arise from the normal ordering between $\mathbf{S}$ and $\tilde{\mathbf{S}}$, since (up to some unimportant factors of $q, t$)
\begin{align}
	\prod_{a = 1}^{n^\text{L}}\mathcal{S}_+(z_a, B_a)_\text{L}\prod_{b = 1}^{n^\text{R}}\mathcal{S}_-(z_b, B_b)_\text{R} = & \ :\prod_{a = 1}^{n^\text{L}}\mathcal{S}_+(z_a, B_a)_\text{L}\prod_{b = 1}^{n^\text{R}}\mathcal{S}_-(z_b, B_b)_\text{R}:  Z_\text{1d chiral} \ .
\end{align}

Finally, summing over all the integers $B_a^{(i)}$, we have the index of an intersecting $U(n^\text{L}) \times U(n^\text{R})$ gauge theory coupled to 3d adjoint chiral multiplets and additional 1d bifundamental chirals given by the matrix element
\begin{align}
	I = \langle w|\sum_{B^{(i)}_a} \oint \prod_{a,i}^{n^{(i)}}\frac{dz_a^{(i)}}{2\pi i z_a^{(i)}}
	\prod_{a = 1}^{n^\text{L}} \mathbf{S}(z_a, B_a)_\text{L}
	\prod_{a = 1}^{n^\text{R}} \mathbf{S}(z_a, B_a)_{(2)} | w \rangle\ ,
\end{align}
where the ket-states $|w \rangle$ and bra-state $\langle w |$ are appropriate Fock states. Finally, it is straightforward to generate pairs of fundamental and anti-fundamental chiral multiplet contributions using a differential operator realization of the Heisenberg operators,
\begin{align}
	& \mathbf{a}_n = \frac{1}{n}(t^{\frac{n}{2}} - t^{- \frac{n}{2}})(p^{\frac{n}{2}} + p^{- \frac{n}{2}}) \frac{d}{d \tau_n} \ ,\qquad
	\mathbf{a}_{- n} = (q^{\frac{n}{2	}} - q^{- \frac{n}{2}}) \tau_n \ ,\\
	& \mathbf{P} = 2 \frac{d}{d\tau_0}, \qquad \mathbf{Q} = \tau_0\ .
\end{align}
A shift
\begin{align}
	\tau_n \to \tau_n + \sum_{i = 1}^N \frac{t_i^{-n} - \tilde t^n_i}{n (1 - q^n)} 
\end{align}
leads to the desired contributions.

We argue that the sum of products of the integrated screening currents above (with appropriate contour) sits in the kernel of both $\mathbf{T}_{q, t}$ and $\mathbf{T}_{\tilde q, \tilde t}$. This can be seen by the following elementary computation done for the modes $\mathbf{T}_m$ of $\mathbf{T}_{q,t}$ (or $\tilde {\mathbf{T}}_m$ for $\mathbf{T}_{\tilde q, \tilde t}$) and $\mathcal{S}_+$,
\begin{align}
	& \ \sum_{B \in \mathbb{Z}}\left[\mathbf{T}_m, q^B\mathbf{S}_+(z q^{\frac{B}{2}})\tilde{\mathbf{S}}'_+(z^{-1}q^{\frac{B}{2}})\right]
	= \sum_{B \in \mathbb{Z}}q^B\left[\mathbf{T}_m, \mathbf{S}_+(z q^{\frac{B}{2}})\right]\tilde{\mathbf{S}}'_+(z^{-1}q^{\frac{B}{2}})\nonumber\\
	= & \ \sum_B \frac{1}{zq^{ - \frac{B}{2} }}\left(\mathcal{O}_m(q zq^{\frac{B}{2}}) - \mathcal{O}_m(zq^{\frac{B}{2}})\right)\tilde{\mathbf{S}}'_+(z^{-1}q^{\frac{B}{2}}) \\
	= & \ \sum_B \frac{1}{(q^{\frac{1}{2}}z)q^{\frac{ - B}{2}}}\mathcal{O}_m(q^{\frac{1}{2}}z)q^{\frac{B}{2}}) \tilde{\mathbf{S}}'_+((z q^{\frac{1}{2}})^{-1}q^{\frac{B}{2}})
	- \sum_B \frac{1}{zq^{ - \frac{B}{2} }}\mathcal{O}_m(zq^{\frac{B}{2}})\tilde{\mathbf{S}}'_+(z^{-1}q^{\frac{B}{2}})\nonumber \\
	\equiv & \ \sum_B \mathcal{O}_m^B(q^{\frac{1}{2}}z) - \mathcal{O}_m^B(z) \ ,\nonumber
\end{align}
which is a total difference, and when integrated with a appropriate contour, the commutator vanishes. In the above we have schematically split the $\mathcal{S}_+$ into a $\mathbf{S}_+$ and a $\tilde{\mathbf{S}}'_+$ piece where the prime indicates that the current lacks the $\mathbf{Q}$ factor, since that factor has been allocated to $\mathbf{S}_+$ which participates in the commutator with $\mathbf{T}_m$ (which commutes with $\tilde {\mathbf{S}}'_+$). The fact that the commutator gives a total difference implies that with appropriate integration contour, the product is annihilated by both the stress tensors.

In the above construction, we effectively glued two modular-doubles built from the two $q$-Virasoro algebras, where each modular-double is engineered to generate the $\mathcal{N} = 2$ superconformal index, and both the modular-double screening charges $\oint \mathcal{S}$ commute with the two $q$-Virasoro stress tensors. We argue that such a construction is in fact the maximal one, in the sense that one cannot glue successively more than two modular doubles of such types while requiring the commutativity between all the screening charges and the $q$-Virasoro stress tensors. For example, consider three $q$-Viraroso algebras $\mathbf{V}_{q_i,t_i}$ generated by the Heisenberg operators $\{\mathbf{a}_{in}, \mathbf{P}_i, \mathbf{Q}_i\}$. One can set $q_2 = q^{-1}_1 \equiv q^{-1}$, $t_2 = t_1^{-1} \equiv t^{-1}$, $q_3 = q_1 = q$, $t_3 = t_1 =t$ and also identify the zero modes $\mathbf{P} = \mathbf{P}_i$, $\mathbf{Q} = \mathbf{Q}_i$ in order to construct modular doubles
\begin{figure}
	\centering
	\includegraphics{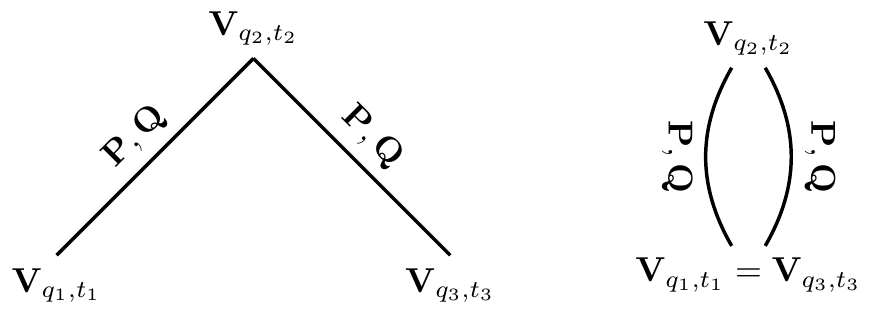}
	\caption{Gluing three $q$-Virasoro algebras would fail the commutativity requirement, unless the third algebra is identified with the first, as shown on the right.}
\end{figure}
\begin{align}
	\mathcal{S}_{12}(z, B) = & \ \exp\left[
	  - \sum \frac{\mathbf{a}_{1n} (z q_1^{\frac{B}{2}})^{-n}}{q_1^{n/2} - q_1^{-n/2}}
    - \sum \frac{\mathbf{a}_{2n} (z^{-1} q_2^{\frac{B}{2}})^{-n}}{q_2^{n/2} - q_2^{-n/2}}
    + \sqrt{\beta} \mathbf{Q}
	\right]q_1^{B\sqrt{\beta}\mathbf{P}}\\
	\mathcal{S}_{23}(z, B) = & \ \exp\left[
	  - \sum \frac{\mathbf{a}_{2n} (z q_1^{\frac{B}{2}})^{-n}}{t_2^{ - n/2} - t_2^{+n/2}}
    - \sum \frac{\mathbf{a}_{3n} (z^{-1} t_2^{\frac{B}{2}})^{-n}}{t_3^{ - n/2} - t_3^{+n/2}}
    - \frac{1}{\sqrt{\beta}} \mathbf{Q}
	\right]t_2^{B \sqrt{\beta}\mathbf{P}} \ .
\end{align}
It is straightforward to observe that $\mathbf{T}_{1m}$ and $\mathbf{T}_{2m}$ commute with $\mathcal{S}_{12}$, but $\mathbf{T}_{1m}$ does not commute with $\mathcal{S}_{23}$, since it commutes only with $\mathbf{a}_{2n}$ and $\mathbf{a}_{3n}$ but not $\mathbf{Q}$. The only remedy one can make is to further identify $\mathbf{a}_{3n}$ with $\mathbf{a}_{1n}$, thus reproducing the construction discussed above.

\section*{Acknowledgments}

The authors would like to thank Yongchao L\"{u}, Fabrizio Nieri and Wolfger Peelaers for suggestions. This work is supported in part by the
National Natural Science Foundation of China (NSFC)
under Grant Nos. 11905301 (YP) and 11875327 (HHZ), the Fundamental Research Funds
for the Central Universities, and the Sun Yat-Sen University
Science Foundation.

\appendix
\section{Special functions}

The $q$-Pochhammer symbol is defined by the (regularized) infinite
\begin{align}
	(z; q) \equiv \prod_{k = 0}^{+\infty} (1- z q^k) \ , \qquad |q| < 1\ .
\end{align}
One can extend the definition to regions with $|q| > 1$ by
\begin{align}
	(z; q) = (z q^{-1};q^{-1})^{-1} \ .
\end{align}

The $q$-Pochhammer symbol has simple and useful series expansions given by
\begin{align}
	(z;q) = \sum_{n = 0}^{+\infty} \frac{(-1)^n q^{\frac{n(n-1)}{2}}}{(q;q)_n} z^n \ ,
	\qquad
	\frac{1}{(z;q)} = \sum_{n = 0}^{+\infty} \frac{z^n}{(q;q)_n} \ .
\end{align}

Similarly, the double $q$-Pochhammer symbol is defined by
\begin{align}
	(z; p, q) \equiv \prod_{k, \ell = 0}^{+\infty}(1 - z p^k q^\ell) \ , \qquad
	|p|, |q| < 1 \ .
\end{align}
To extend beyond the above $p, q$ region, one uses
\begin{align}
	(z; p, q) = \frac{1}{(z p^{-1};p^{-1},q)}
	= \frac{1}{(z q^{-1};p,q^{-1})} = (z p^{-1}q^{-1}; p^{-1}, q^{-1}) \ .
\end{align}

Both $q$-Pochhammer symbols enjoy well-known shift properties. For $m, n \in \mathbb{N}$ ,
\begin{align}
	\frac{(zq^n;q)}{(z;q)} = \frac{1}{\prod_{k = 0}^{n - 1}(1 - zq^k)}\ ,
	\qquad
	\frac{(zq^{ - n};q)}{(z;q)} = \prod_{k = 1}^{n}(1 - zq^{ - k}) \ ,
\end{align}
and also
\begin{align}
  \frac{(z\mathfrak{p}^m \mathfrak{q}^n;\mathfrak{p},\mathfrak{q})}{(z;\mathfrak{p},\mathfrak{q})} = \frac{\prod_{k = 0}^{m-1}\prod_{\ell = 0}^{n - 1}(1 - z \mathfrak{p}^k \mathfrak{q}^\ell)}{\prod_{k = 0}^{m - 1}(z\mathfrak{p}^k;\mathfrak{q}) \prod_{\ell = 0}^{n - 1}(z \mathfrak{q}^\ell;\mathfrak{p})} \ ,
\end{align}
\begin{align}
  \frac{(z\mathfrak{p}^{ - m} \mathfrak{q}^{- n};\mathfrak{p},\mathfrak{q})}{(z; \mathfrak{p},\mathfrak{q})}
  = \prod_{k=1}^{m}(z\mathfrak{p}^{-k};\mathfrak{q})\prod_{\ell=1}^{n}(z\mathfrak{q}^{-\ell};\mathfrak{p}) \prod_{k=1}^{m}\prod_{\ell = 1}^{n}(1 - z\mathfrak{p}^{-k}\mathfrak{q}^{-\ell}) \ .
\end{align}

\section{Instanton partition function\label{app:ipf}}

The $U(N)$ SQCD instanton partition function on $\mathbb{R}^4 \times S^1$ with $\Omega$-deformation paramters $\mathfrak{p} \equiv e^{2\pi i \epsilon_1}$, $\mathfrak{q} \equiv e^{2\pi i \epsilon_2}$ is given by a sum over $N$-tuples of Young diagrams \cite{Nekrasov:2003rj,Nekrasov:2002qd,Tachikawa:2004ur,Sulkowski:2009ne}
\begin{align}
	Z_\text{inst}(Q, k_\text{CS}^\text{5d}; z, \mu, \tilde \mu; \mathfrak{p}, \mathfrak{q}) \equiv \sum_{\vec Y} Q^{|\vec Y|}\mathcal{Z}_\text{CS}(\vec Y; k_\text{CS}^\text{5d},z) \mathcal{Z}_\text{VM}(\vec Y; z) \mathcal{Z}_\text{HM}(\vec Y; z, \mu, \tilde \mu)\ ,
\end{align}
where $Q$ denotes the exponentiated Yang-Mills coupling, and $\mathcal{Z}_\text{CS}$ captues the contribution from the Chern-Simons term \cite{Tachikawa:2004ur},
\begin{align}
	\mathcal{Z}_\text{CS} = \prod_A z_{A}^{ - k^\text{5d}_\text{CS} | Y_A|} \mathfrak{q}^{- \frac{k}{2} || Y_A||^2} \mathfrak{p}^{- \frac{1}{2}|| Y_A^\vee ||^2}, 
\end{align}
For any Young diagram $Y$, we use the symbol $||Y||^2 \equiv \sum_{r = 1} Y_r^2$, and $Y^\vee$ denotes the transposition of the Young diagram $Y$. It can be shown that
\begin{align}
	||Y^\vee||^2 = - \sum_{r = 1}Y_r(1 - 2r) \ .
\end{align}
Although most of the discussions in the main text focus on vanishing Chern-Simons level, however, it is crucial to keep it general in order to completely fix the fugacities relation between the 5d and 3d theories related by Higgsing.

The vector multiplet contribution is given by
\begin{align}
	z_\text{vect} = \prod_{A, B = 1}^N \prod_{r, s = 1}^{\infty}
	  \frac{(\hat{z}_{AB} - \epsilon_1 (s - r + 1) - Y_{Bs} \epsilon_2)_{Y_{Ar}}^{\sinh}}{{(\hat{z}_{AB} - \epsilon_1(s - r + 1) - Y_{Bs}\epsilon_2)_{Y_{Bs}}^{\sinh}}}
	  \frac{(\epsilon_2^{-1}\hat{z}_{AB} - \epsilon_1(s - r) - Y_{Bs}\epsilon_2)_{Y_{Bs}}^{\sinh}}{{(\epsilon_2^{-1}\hat{z}_{AB} - \epsilon_1(s - r) - Y_{Bs}\epsilon_2)_{Y_{Ar}}^{\sinh}}} \ ,
\end{align}
where $z \equiv e^{2\pi i \beta \hat z}$, $\hat z_{AB} = \hat z_{A} - \hat z_B$, and we define the (regularized) infinite product
\begin{align}
	(x)^{\sinh}_m \equiv \prod_{k = 0}^{m - 1} 2\sinh\pi i \beta (x + k \epsilon_2) \ .
\end{align}
The fundamental and anti-fundamental hypermultiplets contribute
\begin{align}
	z_\text{(a)fund}(\vec Y; z, \mu^\epsilon_I) = & \ \prod_{A = 1}^N\prod_{I = 1}^{N} \prod_{r = 1}^{+\infty}(\hat{z}_A - \mu_I^\epsilon + \epsilon_1 r + \epsilon_2)_{Y_{Ar}}^{\sinh} \nonumber \\
  = & \ \prod_{A = 1}^N \prod_{I = 1}^{N}\prod_{r = 1}^{+\infty}\prod_{s = 1}^{Y_{Ar}}2 \sinh \pi i \beta( \hat z_A - \hat \mu^\epsilon_I + r \epsilon_1 + s \epsilon_2)\ ,
\end{align}

When $z_A$ are specified to the special values $z_A = \mu_A \mathfrak{p}^{-n^\text{L}_A - \frac{1}{2}} \mathfrak{q}^{- n^\text{R}_A - \frac{1}{2}}$, each summand corresponding to a tuple $\vec Y$ factorizes. In particular, when $\vec Y$ is a tuple of only large Young diagrams with respect to the forbidden boxes $\{(n^\text{L}_A + 1, n^\text{R} + 1)\}_{A = 1}^N$, we can encode $\vec Y$ into the sequences $\{\mathfrak{m}^\text{L,R}_{A \mu}\}$, and we have
\begin{align}
	\mathcal{Z}_\text{afund}(\vec Y) \to & Z^{\vec n^\text{L}}_\text{vortex-afund}(\mathfrak{m},q)_\text{L}Z^{\vec n^\text{R}}_\text{vortex-afund}(\mathfrak{m},q)_\text{R} (Z^{\vec n^\text{L} \vec n^\text{R}}_\text{afund-extra})^{-1} \\
  \mathcal{Z}_\text{VF}(\vec Y) \to & Z^{\vec n^\text{L}}_\text{vortex-fund-adj}(\mathfrak{m},q)_\text{L}Z^{\vec n^\text{R}}_\text{vortex-fund-adj}(\mathfrak{m},q)_\text{R} Z^{\vec n^\text{L}, \vec n^\text{R}}_\text{intersection}(\mathfrak{m}^\text{L}, \mathfrak{m}^\text{R}) (Z^{\vec n^\text{L} \vec n^\text{R}}_\text{VF-extra})^{-1}\  .
\end{align}
Here the factors $Z_\text{vortex-afund}$ and $Z_\text{vortex-fund-adj}$ are factors in the vortex partition function summand (\ref{vpf}) that we review in the next appendix. The parameters in the 5d and the 3d theories are identified by
\begin{align}
	& (t_i \tau)_\text{L} = \mu_i \mathfrak{p}^{-1}\mathfrak{q}^{ - 1/2} \ ,\qquad
	(\tilde t_i \tau)_\text{L} = \tilde \mu_i^{-1}\mathfrak{q}^{1/2} \ ,\qquad
	v_\text{L} = \mathfrak{p}^{-1}, \qquad q_\text{L} = \mathfrak{q}\ , \\
	& (t_i \tau)_\text{R} = \mu_i \mathfrak{q}^{-1}\mathfrak{p}^{ - 1/2} \ ,\qquad
	(\tilde t_i \tau)_\text{R} = \tilde \mu_i^{-1}\mathfrak{p}^{1/2} \ ,\qquad
	v_\text{R} = \mathfrak{q}^{-1}, \qquad q_\text{R} = \mathfrak{p} \ .
\end{align}

The $Q$ factor and the Chern-Simons term also happily factorize when $z$ is specialized. It is easy to see that the former indeed factorizes into $Q^{|\vec Y|} = Q^{\sum_{A = 1}^N |\mathfrak{m}^\text{L}_A|} Q^{\sum_{A = 1}^N |\mathfrak{m}^\text{R}_A|} Q^{\sum_{A = 1}^N n^\text{L}_A n^\text{R}_A}$. The latter requires a bit of straightforward computation. At the special $z$, the Chern-Simons factor reads
\begin{align}
	\prod_{A = 1}^N
	\mu_A^{- k^\text{5d}_\text{CS} |Y_A|}
	\mathfrak{p}^{
	  (n^\text{L}_A + \frac{1}{2})k^\text{5d}_\text{CS}\sum_{s = 1} Y^\vee_{As}
	  - \frac{k^\text{5d}_\text{CS}}{2} \sum_{s = 1}(Y^\vee_{As})^2
	}
	\mathfrak{q}^{
	  (n^\text{R}_A + \frac{1}{2})k^\text{5d}_\text{CS}\sum_{r = 1} Y_{Ar}
	  - \frac{k^\text{5d}_\text{CS}}{2} \sum_{r = 1}(Y_{Ar})^2
	} \ .
\end{align}
One can rewrite $Y_{A(n^\text{L}_A - \mu)} = n^\text{R}_A + \mathfrak{m}^\text{L}_{A\mu}$, for $ \mu = 0, \ldots, n^\text{L}_A - 1$, $(Y^\vee_{A})_{n^\text{R}_A - \nu} = n^\text{L}_A + \mathfrak{m}^\text{R}_{A\mu}$, for $ \nu = 0, \ldots, n^\text{R}_A - 1$, and
\begin{align}
  \sum_{r = n_A^\text{L} + 1}^{+\infty}(Y_{Ar})^2
  = & \ ||((Y_A^\text{R})^\vee )^\vee|| ^2
  = - \sum_{s = 1}^{n_A^\text{R}} (Y_A^\text{R})^\vee_s (1 - 2s) \ , \\
  \sum_{r = n_A^\text{L} + 1}^{+\infty}(Y^\vee_{A})_s^2
  = & \ ||(Y_A^\text{L})^\vee || ^2
  = - \sum_{r = 1}^{n_A^\text{L}} Y_{Ar}^\text{L} (1 - 2r) \ ,
\end{align}
where again $Y^\text{L}_{Ar} = \mathfrak{m}^\text{L}_{A (n^\text{L}_A - r)}$, $Y^\text{R}_{As} = \mathfrak{m}^\text{R}_{A (n^\text{R}_A - s)}$ for $r = 1, \ldots, n^\text{L}_A$, $s = 1, \ldots, n^\text{R}_A$. In the end, the Chern-Simons factor factorizes
\begin{align}\label{5d-3d-mass-relation-app}
	= & \ \prod_{A = 1}^{N}(\mu_A \mathfrak{p}^{ - \frac{n_A^\text{L} + 1}{2}} \mathfrak{q}^{- \frac{n_A^\text{R} + 1}{2}})^{- k^\text{5d}_\text{CS}n^\text{L}_A n^\text{R}_A} \\
	& \ \times \prod_{\mu = 0}^{n^\text{L}_A - 1} \left( \mu_A \mathfrak{q}^{-1/2}\mathfrak{p}^{-1} \mathfrak{p}^{-\mu}\mathfrak{q}^{\frac{1}{2} \mathfrak{m}^\text{L}_{A\mu} }\right)^{-k_\text{CS}^\text{5d} \mathfrak{m}^\text{L}_{A\mu}}
	\prod_{\nu = 0}^{n^\text{R}_A - 1} \left( \mu_A \mathfrak{p}^{-1/2}\mathfrak{q}^{-1} \mathfrak{q}^{-\nu}\mathfrak{p}^{\frac{1}{2} \mathfrak{m}^\text{R}_{A\nu} }\right)^{-k_\text{CS}^\text{5d} \mathfrak{m}^\text{R}_{A\nu}} \ . \nonumber
\end{align}
Let us pause and comment on the role of the Chern-Simons factor in parameter identification. Without the Chern-Simons factor, the vortex partition function depends only on the ratios, e.g. $t_it_j^{-1}$, between flavor fugacities, and therefore the identification of the vortex partition function at vanishing Chern-Simons level only fixes the relations between the 5d fugacity ratios and 3d fugacity ratios. The Chern-Simons term however provide additional constraints which lead to the final complete identification (\ref{5d-3d-mass-relation-app}).

\section{Factorization of 3d index}

Here we collect some detail on the factorization of the index of a 3d $\mathcal{N} = 2$ $U(n)$ gauge theory with Chern-Simons level $k_\text{CS}$ coupled to an adjoint, $n_\text{f}$ fundamental and $n_\text{af}$ antifundamental chiral multiplets. Part of the following computation follows that in \cite{Hwang:2015wna}, while towards the end the result is reorganized following \cite{Pan:2016fbl}. 

The index can be computed in terms of a contour integral
\begin{align}
  I = \ \frac{1}{n!} \sum_{\vec B \in \mathbb{Z}^n} \oint _{|z_a| = 1} \prod_{a = 1}^n  \frac{dz_a}{2\pi i z_a} & \prod_{a = 1}^n(-z_a)^{-k_\text{CS} B_a} (-w)^{B_a} \prod_{a \ne b} q^{ - \frac{|B_a - B_b|}{4}} (1 - z_a z^{-1}_b q^{\frac{|B_a - B_b|}{2}}) \nonumber \\
  & \ \times \prod_{a,b=1}^n (q^{\frac{ - 1}{2}} v)^{- \frac{|B_a - B_b|}{2}} \frac{(z_a z_b^{-1}v^{-1} q^{\frac{2 + |B_a - B_b|}{2}};q)}{(z_a^{-1} z_b v  q^{\frac{|B_a - B_b|}{2}};q)}\\
  & \ \times \prod_{i = 1}^{n_\text{f}}\prod_{a = 1}^n(q^{\frac{ - 1}{2}} ( - z_a^{-1}) (t_i \tau))^{- \frac{|B_a|}{2}} \frac{(z_a (t_i \tau)^{-1} q^{\frac{2 + |B_a|}{2}};q)}{(z_a^{-1} (t_i \tau)  q^{\frac{|B_a|}{2}};q)} \nonumber \\
  & \ \times \prod_{i = 1}^{n_\text{af}}\prod_{a=1}^n (q^{\frac{ - 1}{2}}( - z_a) (\tilde t_i \tau))^{- \frac{|B_a|}{2}} \frac{(z_a^{-1} (\tilde t_i \tau)^{-1} q^{\frac{2 + |B_a|}{2}};q)}{(z_a (\tilde t_i \tau)  q^{\frac{|B_a|}{2}};q)} \ . \nonumber
\end{align}

The factorization computation begins with identifying a set of poles coming from the fundamental one-loop contribution. They are labeled by the partitions $\{n_i, i = 1,\ldots, n_\text{f}\}$ of $n$ (which labels Higgs vacua of the gauge theory) and two sets of non-decreasing sequence of natural numbers $\mathfrak{m}_{i\mu}, \overline{\mathfrak{m}}_{i\mu}$ where $\mu = 0, 1, \ldots, n_i - 1$. Explicitly, they are given by
\begin{align}
	z_{i\mu}^+ \equiv z_{i\mu}q^{\frac{B_{i\mu}}{2}} = t_i \tau v^{\mu} q^{\mathfrak{m}_{i \mu}},
	\qquad
	z_{i\mu}^- \equiv z_{i\mu}q^{ - \frac{B_{i\mu}}{2}} = t_i \tau v^{\mu} q^{\overline{\mathfrak{m}}_{i \mu}} \ ,
\end{align}
where $B$ and $\mathfrak{m}, \overline{\mathfrak{m}}$ are related by
\begin{align}
	B_{i \mu} = \mathfrak{m}_{i \mu} - \overline{\mathfrak{m}}_{i \mu} \ .
\end{align}

To compute the residue at these poles, it is more convenient to rewrite the integrand using the identities for any integer $B$
\begin{align}
   \frac{(x^{-1}q^{1 + \frac{1}{2}|B|};q)}{(x q^{\frac{1}{2}|B|};q)}
   = (- x q^{- \frac{1}{2}})^{\frac{1}{2}(|B| \mp B)}\frac{(x^{-1}q^{1 \pm \frac{1}{2}B})}{(x q^{\pm \frac{1}{2}B};q)} \ ,
\end{align}
and
\begin{align}
	\left(1 - z q^{|B|}\right)\left(1 - z^{-1} q^{|B|}\right)
	= - q^{|B|}[ (zq^B)^{1/2} - (zq^B)^{-1/2}] [ (zq^{ - B})^{1/2} - (zq^{ - B})^{-1/2}] \ .
\end{align}

As a result, we remove almost all the absolute values in the integral and we have
\begin{align}
  I = \ \frac{1}{n!} \sum_{\vec B \in \mathbb{Z}^n} \oint _{|z_a| = 1} &  \prod_{a = 1}^n  \frac{dz_a}{2\pi i z_a} (-z_a)^{-k_\text{CS} B_a} (- w)^{B_a} \prod_{a < b} \left[
      \left(\frac{z_a^+}{z_b^+}\right)^{1/2} -
      \left(\frac{z_a^+}{z_b^+}\right)^{-1/2}
    \right]
    \left[ z^+ \to z^-\right] \nonumber\\
  & \ \times \prod_{a,b=1}^n (-1)^{\frac{1}{2}|B_a - B_b|}\left(q^{\frac{ - 1}{2}} v\frac{z_b}{z_a}\right)^{\frac{B_a - B_b}{2}}
      \frac{( (z^-_a/z^-_b)v^{-1} q;q)}{((z^+_b/z_a^+)^{-1} v;q)}\nonumber\\
  & \ \times \prod_{i = 1}^{n_\text{f}}\prod_{a = 1}^n
    (q^{\frac{ - 1}{2}} ( - z_a^{-1}) (t_i \tau))^{+ \frac{B_a}{2}}
    \frac{(z^-_a (t_i \tau)^{-1} q;q)}{((z^+_a)^{-1} (t_i \tau) ;q)}\nonumber\\
  & \ \times \prod_{i = 1}^{n_\text{af}}\prod_{a=1}^n
    (q^{\frac{ - 1}{2}}( - z_a) (\tilde t_i \tau))^{- \frac{B_a}{2}}
    \frac{((z_a^-)^{-1} (\tilde t_i \tau)^{-1} q ;q)}{(z_a^+ (\tilde t_i \tau)  ;q)} \ ,\nonumber
	\end{align}
where again $z_a^\pm \equiv z_a q^{\pm \frac{B_a}{2}}$.

It is starightforward to compute the residue at the above listed poles. Corresponding to a given partition $\vec n \equiv \{n_i\}$ of $n$, by standard arguments the sum over poles joins with the sum over magnetic fluxes $\vec B$ to form a double sum over sequences denoted as $\mathfrak{m}, \overline{\mathfrak{m} } \ge 0$. Using the shift properties of the $q$-Pochhammer symbols, one can separate factors independent of $\mathfrak{m}, \overline{\mathfrak{m}}$ and group them into the Higgs-branch-localized one-loop factors $Z_\text{1-loop}^{\vec n}$, while the remaining factors that do depend on $\mathfrak{m}, \overline{\mathfrak{m}}$ will be grouped into the vortex partition function. Explicitly
\begin{align}
	Z_\text{1-loop}^{\vec n} = \prod_{j = 1 }^{n_\text{af}}\prod_{i = 1}^{n_\text{f}}\prod_{\mu = 0}^{n_i - 1}\frac{((t_i \tilde t_j \tau^2 v^\mu)^{-1} q;q)}{(t_i \tilde t_j \tau^2 v^\mu; q)}\prod_{i, j = 1}^{n_\text{f}}\prod_{\mu = 0}^{n_i - 1} \frac{(t_{ij}v^{- n_j + \mu}q;q)}{(t_{ij}^{-1}v^{n_j - \mu};q)} \ ,
\end{align}
where the first factor obviously comes from the anti-fundamental chiral one-loop, and the second factor comes from the fundamental and adjoint chirals\footnote{To obtain this factor, it is convenient to apply the following identity with $f(x) \equiv (x^{-1}q;q)(x;q)^{-1}$
\begin{align}
  & \ \prod_{i,j = 1}^{n_\text{f}}\prod_{\mu = 0}^{n_i - 1}f(t_{ij}^{-1}v^{-\mu})
  \prod_{i, j = 1}^{n_\text{f}}\prod_{\mu = 0}^{n_i - 1}\prod_{\nu = 0}^{n_j - 1}f(t_{ij}^{-1}v^{\nu - \mu + 1}) \nonumber\\
  =
  & \ f(0)^n
  \prod_{i,j = 1}^{n_\text{f}}\prod_{\mu = 0}^{n_i - 1}f(t_{ij}^{-1} v^{n_j - \mu})
  \prod_{i \ne j|n_i \ge n_j} \prod_{\mu = 0}^{n_i - 1}\prod_{\nu = 0}^{n_j - 1}f(t_{ij}^{-1}v^{-(\mu - \nu)})f(t_{ij}v^{\mu - \nu})
  \prod_{i = 1}^{n_\text{f}} \prod_{\substack{\mu, \nu = 0\\\mu > \nu }}^{n_i - 1} f(v^{\nu - \mu})f(v^{ - \nu + \mu})\ .
\end{align}
Here the first factor is related to the residues $\left(\operatorname{Res}\frac{(q;q)}{(z^{-1};q)}\right)^n$, while the second factor will be the one that is grouped into the Higgs-branch-localized one-loop. The remaining factors will be sent into the vortex piece; note that
\begin{align}
	f(x)f(x^{-1}) = \frac{1}{(1- x)(1 - x^{-1})} = \frac{1}{- (x^{1/2} - x^{-1/2})(x^{1/2} - x^{-1/2})} \ .
\end{align}
}. Also, $t_{ij}$ abbreviates $t_i t_j^{-1}$.

The remaining factors go into the vortex partition function which factorizes into those containing $\mathfrak{m}$ and those containing $\overline{\mathfrak{m}}$. The former reads
\begin{align}
  = & \ \prod_{i = 1}^{n_\text{f}}\prod_{\mu = 0}^{n_i - 1}
  (-t_i \tau v^\mu q^{\frac{1}{2} \mathfrak{m}_{i \mu}})^{\frac{k_\text{CS}}{2}\mathfrak{m}_{i\mu}}(-w)^{\mathfrak{m}_{i\mu}} \\
  & \times \prod_{i,j = 1}^{n_\text{f}}\prod_{\mu, \nu}(q^{-\frac{1}{2}} t_{ij}^{-1}v^{\nu - \mu + 1})^{\frac{1}{2}(\mathfrak{m}_{i\mu} - \mathfrak{m}_{j\nu})}
  q^{- \frac{1}{4}(\mathfrak{m}_{i \mu} - \mathfrak{m}_{j\nu})^2} \\
  & \times \prod_{i,j=1}^{n_\text{f}}\prod_{\mu = 0}^{n_i}(-t_{ij}^{-1}v^{ - \mu} q^{- \frac{1}{2}})^{\frac{\mathfrak{m}_{i\mu}}{2}} q^{ - \frac{\mathfrak{m}^2_{i\mu}}{4}}\prod_{j = 1}^{n_\text{af}}\prod_{i = 1}^{n_\text{f}}\prod_{\mu = 0}^{n_i - 1} (- q^{-\frac{1}{2}}t_i \tilde t_j \tau^2 v^{\mu })^{- \frac{\mathfrak{m}_{i \mu} }{2}}q^{- \frac{\mathfrak{m^2_{i \mu}}}{4}}\\
	& \times \prod_{i = 1}^{n_\text{f}}\prod_{\mu = 0}^{n_i - 1}
	    \frac{
	    \prod_{j = 1}^{n_\text{af}}\prod_{k = 0}^{\mathfrak{m}_{i \mu} - 1} 1 - t_i \tilde t_j \tau^2 v^\mu q^k
	    }{
	    \prod_{j = 1}^{n_\text{f}}\prod_{k = 1}^{\mathfrak{m}_{i \mu}} 1 - t_{ij}^{-1} v^{-\mu}q^{-k}
	    }
	  \prod_{(i, \mu)}\prod_{(j, \nu)} \left[
	    \frac{
	      \prod_{k = 0}^{\mathfrak{m}_{j \nu} - 1} (1 - t_{ij}^{-1}v^{\nu - \mu + 1} q^{k - \mathfrak{m}_{i \mu}})
	    }{
	      \prod_{k = 1}^{\mathfrak{m}_{i \mu}}(1 - t_{ij}^{-1}v^{\nu - \mu + 1} q^{ - k})
	    }
	  \right] \\
	 & \times \prod_{(i, \mu) > (j, \nu)}(-1)^{\mathfrak{m}_{i \mu} - \mathfrak{m}_{j \nu}}\prod_{a< b} \frac{-\sinh\pi i \beta(\hat t_{ab} + \hat \nu(\mu_a - \mu_b) + \hat q(\mathfrak{m}_a - \mathfrak{m}_b)) }{ - \sinh\pi i\beta(\hat t_{ab} + \hat \nu(\mu_a - \mu_b))}\ ,\nonumber
\end{align}
where we define the hatted variables uniformly by $x = e^{2\pi i \beta \hat x}$. Some more massaging turns this expression into a more recognizable form
\begin{align}\label{vpf}
	& \ \prod_{i = 1}^{n_\text{f}}\prod_{\mu = 0}^{n_i - 1}
  (-t_i \tau v^\mu q^{\frac{\mathfrak{m}_{i\mu}}{2}})^{- k_\text{CS}\mathfrak{m}_{i\mu}}(-w)^{\mathfrak{m}_{i\mu}}
  \prod_{j = 1}^{n_\text{af}}\prod_{i = 1}^{n_\text{f}}\prod_{\mu = 0}^{n_i - 1}\prod_{k = 0}^{\mathfrak{m}_{i\mu} - 1} \frac{1
  }{
    2\sinh\pi i \beta(\hat t_i + \hat {\tilde t}_j + \mu \hat v + (\mathfrak{m}_{i\mu} - k)\hat q) 
  }\nonumber
  \\
  & \ \times \prod_{i,j}^{n_\text{f}}\Bigg[
    \prod_{\mu = 0}^{n_i - 1}\prod_{\nu = 0}^{n_j - 1} \frac{1}{\prod_{k = 0 }^{\mathfrak{m}_{i\mu} - \mathfrak{m}_{i, \mu - 1}} 2\sinh \pi i \beta(\hat t_{ij} + (\mu - \nu) \hat v + (\mathfrak{m}_{i \mu} - \mathfrak{m}_{j\nu} - k)\hat q)}\\
  & \ \qquad \qquad\times \prod_{\mu = 0}^{n_i - 1} \frac{\prod_{k = 0}^{\mathfrak{m}_{j,n_j-1} - 1}2\sinh \pi i \beta(- \hat t_{ij} + (n_j - \mu)\hat v - (\mathfrak{m}_{i\mu} - \mathfrak{m}_{j, n_j - 1} + 1 + k)\hat q)
  }{
   \prod_{k = 0}^{\mathfrak{m}_{i\mu} - 1} 2\sinh \pi i \beta (- \hat t_{ij} - (\mu - n_j)\hat v - (k + 1)\hat q)
  } \Bigg] \ . \nonumber
\end{align}
This expression makes up the summand $Z_\text{vortex}^{\vec n}(\mathfrak{m}; k_\text{CS}, w;t, \tilde t, v, \tau; q)$ of the full vortex partition function
\begin{align}
  Z_\text{vortex}^{\vec n}(k_\text{CS}, w; t, \tilde t, v, \tau; q) = \sum_\mathfrak{m} Z_\text{vortex}^{\vec n}(\mathfrak{m}; k_\text{CS}, w;t, \tilde t, v, \tau; q) \ .
\end{align}

Finally, the index is now written in a factorized form
\begin{align}
  I = \sum_{\vec n} Z^{\vec n}_\text{1-loop}
  Z^{\vec n}_\text{vortex} (k_\text{CS}, w; t, \tilde t, v, \tau; q)
  Z^{\vec n}_\text{vortex} (k_\text{CS}, w^{-1}; t^{-1}, \tilde t^{-1}, v^{-1}, \tau^{-1}; q^{-1}) \ .
\end{align}

\clearpage

{
\bibliographystyle{utphys}
\bibliography{ref}

\providecommand{\href}[2]{#2}\begingroup\raggedright\begin{thebibliography}{10}

\bibitem{Pestun:2007rz}
V.~Pestun, ``{Localization of gauge theory on a four-sphere and supersymmetric
  Wilson loops},'' \href{http://dx.doi.org/10.1007/s00220-012-1485-0}{{\em
  Commun. Math. Phys.} {\bf 313} (2012)  71--129},
  \href{http://arxiv.org/abs/0712.2824}{{\tt arXiv:0712.2824 [hep-th]}}.

\bibitem{Kimura:2018axa}
T.~Kimura, J.~Nian, and P.~Zhao, ``{Partition functions of $\mathcal{N}=1$
  gauge theories on $S^2 \times \mathbb{R}^2_\epsilon$ and duality},''
  \href{http://dx.doi.org/10.1142/S0217751X20502073}{{\em Int. J. Mod. Phys. A}
  {\bf 35} (2020) no.~33, 2050207}, \href{http://arxiv.org/abs/1812.11188}{{\tt
  arXiv:1812.11188 [hep-th]}}.

\bibitem{Longhi:2019hdh}
P.~Longhi, F.~Nieri, and A.~Pittelli, ``{Localization of 4d $\mathcal{N}=1$
  theories on $\mathbb{D}^2\times \mathbb{T}^2$},''
  \href{http://dx.doi.org/10.1007/JHEP12(2019)147}{{\em JHEP} {\bf 12} (2019)
  147}, \href{http://arxiv.org/abs/1906.02051}{{\tt arXiv:1906.02051
  [hep-th]}}.

\bibitem{Sugiyama:2020uqh}
K.~Sugiyama and Y.~Yoshida, ``{Supersymmetric indices on $I \times T^2$,
  elliptic genera and dualities with boundaries},''
  \href{http://dx.doi.org/10.1016/j.nuclphysb.2020.115168}{{\em Nucl. Phys. B}
  {\bf 960} (2020)  115168}, \href{http://arxiv.org/abs/2007.07664}{{\tt
  arXiv:2007.07664 [hep-th]}}.

\bibitem{Yoshida:2014ssa}
Y.~Yoshida and K.~Sugiyama, ``{Localization of 3d $\mathcal{N}=2$
  Supersymmetric Theories on $S^1 \times D^2$},''
  \href{http://dx.doi.org/10.1093/ptep/ptaa136}{{\em PTEP} {\bf 2020} (2020)
  11}, \href{http://arxiv.org/abs/1409.6713}{{\tt arXiv:1409.6713 [hep-th]}}.

\bibitem{Nieri:2015yia}
F.~Nieri and S.~Pasquetti, ``{Factorisation and holomorphic blocks in 4d},''
  \href{http://dx.doi.org/10.1007/JHEP11(2015)155}{{\em JHEP} {\bf 11} (2015)
  155}, \href{http://arxiv.org/abs/1507.00261}{{\tt arXiv:1507.00261
  [hep-th]}}.

\bibitem{Pasquetti:2011fj}
S.~Pasquetti, ``{Factorisation of N = 2 Theories on the Squashed 3-Sphere},''
  \href{http://dx.doi.org/10.1007/JHEP04(2012)120}{{\em JHEP} {\bf 04} (2012)
  120}, \href{http://arxiv.org/abs/1111.6905}{{\tt arXiv:1111.6905 [hep-th]}}.

\bibitem{Hwang:2015wna}
C.~Hwang and J.~Park, ``{Factorization of the 3d superconformal index with an
  adjoint matter},'' \href{http://dx.doi.org/10.1007/JHEP11(2015)028}{{\em
  JHEP} {\bf 11} (2015)  028}, \href{http://arxiv.org/abs/1506.03951}{{\tt
  arXiv:1506.03951 [hep-th]}}.

\bibitem{Benini:2012ui}
F.~Benini and S.~Cremonesi, ``{Partition Functions of ${\mathcal{N}=(2,2)}$
  Gauge Theories on S$^{2}$ and Vortices},''
  \href{http://dx.doi.org/10.1007/s00220-014-2112-z}{{\em Commun. Math. Phys.}
  {\bf 334} (2015) no.~3, 1483--1527},
  \href{http://arxiv.org/abs/1206.2356}{{\tt arXiv:1206.2356 [hep-th]}}.

\bibitem{Doroud:2012xw}
N.~Doroud, J.~Gomis, B.~Le~Floch, and S.~Lee, ``{Exact Results in D=2
  Supersymmetric Gauge Theories},''
  \href{http://dx.doi.org/10.1007/JHEP05(2013)093}{{\em JHEP} {\bf 05} (2013)
  093}, \href{http://arxiv.org/abs/1206.2606}{{\tt arXiv:1206.2606 [hep-th]}}.

\bibitem{Hama:2010av}
N.~Hama, K.~Hosomichi, and S.~Lee, ``{Notes on SUSY Gauge Theories on
  Three-Sphere},'' \href{http://dx.doi.org/10.1007/JHEP03(2011)127}{{\em JHEP}
  {\bf 03} (2011)  127}, \href{http://arxiv.org/abs/1012.3512}{{\tt
  arXiv:1012.3512 [hep-th]}}.

\bibitem{Hama:2011ea}
N.~Hama, K.~Hosomichi, and S.~Lee, ``{SUSY Gauge Theories on Squashed
  Three-Spheres},'' \href{http://dx.doi.org/10.1007/JHEP05(2011)014}{{\em JHEP}
  {\bf 05} (2011)  014}, \href{http://arxiv.org/abs/1102.4716}{{\tt
  arXiv:1102.4716 [hep-th]}}.

\bibitem{Fujitsuka:2013fga}
M.~Fujitsuka, M.~Honda, and Y.~Yoshida, ``{Higgs branch localization of 3d
  \ensuremath{\mathcal{N}} = 2 theories},''
  \href{http://dx.doi.org/10.1093/ptep/ptu158}{{\em PTEP} {\bf 2014} (2014)
  no.~12, 123B02}, \href{http://arxiv.org/abs/1312.3627}{{\tt arXiv:1312.3627
  [hep-th]}}.

\bibitem{Benvenuti:2011ga}
S.~Benvenuti and S.~Pasquetti, ``{3D-partition functions on the sphere: exact
  evaluation and mirror symmetry},''
  \href{http://dx.doi.org/10.1007/JHEP05(2012)099}{{\em JHEP} {\bf 05} (2012)
  099}, \href{http://arxiv.org/abs/1105.2551}{{\tt arXiv:1105.2551 [hep-th]}}.

\bibitem{Benini:2013yva}
F.~Benini and W.~Peelaers, ``{Higgs branch localization in three dimensions},''
  \href{http://dx.doi.org/10.1007/JHEP05(2014)030}{{\em JHEP} {\bf 05} (2014)
  030}, \href{http://arxiv.org/abs/1312.6078}{{\tt arXiv:1312.6078 [hep-th]}}.

\bibitem{Hama:2012bg}
N.~Hama and K.~Hosomichi, ``{Seiberg-Witten Theories on Ellipsoids},''
  \href{http://dx.doi.org/10.1007/JHEP09(2012)033}{{\em JHEP} {\bf 09} (2012)
  033}, \href{http://arxiv.org/abs/1206.6359}{{\tt arXiv:1206.6359 [hep-th]}}.
  [Addendum: JHEP 10, 051 (2012)].

\bibitem{Kapustin:2011jm}
A.~Kapustin and B.~Willett, ``{Generalized Superconformal Index for Three
  Dimensional Field Theories},'' \href{http://arxiv.org/abs/1106.2484}{{\tt
  arXiv:1106.2484 [hep-th]}}.

\bibitem{Pan:2015hza}
Y.~Pan and W.~Peelaers, ``{Ellipsoid partition function from Seiberg-Witten
  monopoles},'' \href{http://dx.doi.org/10.1007/JHEP10(2015)183}{{\em JHEP}
  {\bf 10} (2015)  183}, \href{http://arxiv.org/abs/1508.07329}{{\tt
  arXiv:1508.07329 [hep-th]}}.

\bibitem{Chen:2015fta}
H.-Y. Chen and T.-H. Tsai, ``{On Higgs branch localization of
  Seiberg\textendash{}Witten theories on an ellipsoid},''
  \href{http://dx.doi.org/10.1093/ptep/ptv188}{{\em PTEP} {\bf 2016} (2016)
  no.~1, 013B09}, \href{http://arxiv.org/abs/1506.04390}{{\tt arXiv:1506.04390
  [hep-th]}}.

\bibitem{Kallen:2012va}
J.~K\"all\'en, J.~Qiu, and M.~Zabzine, ``{The perturbative partition function
  of supersymmetric 5D Yang-Mills theory with matter on the five-sphere},''
  \href{http://dx.doi.org/10.1007/JHEP08(2012)157}{{\em JHEP} {\bf 08} (2012)
  157}, \href{http://arxiv.org/abs/1206.6008}{{\tt arXiv:1206.6008 [hep-th]}}.

\bibitem{Kallen:2012cs}
J.~K\"all\'en and M.~Zabzine, ``{Twisted supersymmetric 5D Yang-Mills theory
  and contact geometry},''
  \href{http://dx.doi.org/10.1007/JHEP05(2012)125}{{\em JHEP} {\bf 05} (2012)
  125}, \href{http://arxiv.org/abs/1202.1956}{{\tt arXiv:1202.1956 [hep-th]}}.

\bibitem{Minahan:2015jta}
J.~A. Minahan and M.~Zabzine, ``{Gauge theories with 16 supersymmetries on
  spheres},'' \href{http://dx.doi.org/10.1007/JHEP03(2015)155}{{\em JHEP} {\bf
  03} (2015)  155}, \href{http://arxiv.org/abs/1502.07154}{{\tt
  arXiv:1502.07154 [hep-th]}}.

\bibitem{Dedushenko:2016jxl}
M.~Dedushenko, S.~S. Pufu, and R.~Yacoby, ``{A one-dimensional theory for Higgs
  branch operators},'' \href{http://dx.doi.org/10.1007/JHEP03(2018)138}{{\em
  JHEP} {\bf 03} (2018)  138}, \href{http://arxiv.org/abs/1610.00740}{{\tt
  arXiv:1610.00740 [hep-th]}}.

\bibitem{Dedushenko:2017avn}
M.~Dedushenko, Y.~Fan, S.~S. Pufu, and R.~Yacoby, ``{Coulomb Branch Operators
  and Mirror Symmetry in Three Dimensions},''
  \href{http://dx.doi.org/10.1007/JHEP04(2018)037}{{\em JHEP} {\bf 04} (2018)
  037}, \href{http://arxiv.org/abs/1712.09384}{{\tt arXiv:1712.09384
  [hep-th]}}.

\bibitem{Pan:2017zie}
Y.~Pan and W.~Peelaers, ``{Chiral Algebras, Localization and Surface
  Defects},'' \href{http://dx.doi.org/10.1007/JHEP02(2018)138}{{\em JHEP} {\bf
  02} (2018)  138}, \href{http://arxiv.org/abs/1710.04306}{{\tt
  arXiv:1710.04306 [hep-th]}}.

\bibitem{Pan:2019bor}
Y.~Pan and W.~Peelaers, ``{Schur correlation functions on $S^3\times S^1$},''
  \href{http://dx.doi.org/10.1007/JHEP07(2019)013}{{\em JHEP} {\bf 07} (2019)
  013}, \href{http://arxiv.org/abs/1903.03623}{{\tt arXiv:1903.03623
  [hep-th]}}.

\bibitem{Dedushenko:2019yiw}
M.~Dedushenko and M.~Fluder, ``{Chiral Algebra, Localization, Modularity,
  Surface defects, And All That},''
  \href{http://dx.doi.org/10.1063/5.0002661}{{\em J. Math. Phys.} {\bf 61}
  (2020) no.~9, 092302}, \href{http://arxiv.org/abs/1904.02704}{{\tt
  arXiv:1904.02704 [hep-th]}}.

\bibitem{Pan:2019shz}
Y.~Pan and W.~Peelaers, ``{Deformation quantizations from vertex operator
  algebras},'' \href{http://dx.doi.org/10.1007/JHEP06(2020)127}{{\em JHEP} {\bf
  06} (2020)  127}, \href{http://arxiv.org/abs/1911.09631}{{\tt
  arXiv:1911.09631 [hep-th]}}.

\bibitem{Dedushenko:2019mnd}
M.~Dedushenko and Y.~Wang, ``{4d/2d $\rightarrow $ 3d/1d: A song of protected
  operator algebras},'' \href{http://arxiv.org/abs/1912.01006}{{\tt
  arXiv:1912.01006 [hep-th]}}.

\bibitem{Panerai:2020boq}
R.~Panerai, A.~Pittelli, and K.~Polydorou, ``{Topological Correlators and
  Surface Defects from Equivariant Cohomology},''
  \href{http://dx.doi.org/10.1007/JHEP09(2020)185}{{\em JHEP} {\bf 09} (2020)
  185}, \href{http://arxiv.org/abs/2006.06692}{{\tt arXiv:2006.06692
  [hep-th]}}.

\bibitem{Oh:2019bgz}
J.~Oh and J.~Yagi, ``{Chiral algebras from $\Omega$-deformation},''
  \href{http://dx.doi.org/10.1007/JHEP08(2019)143}{{\em JHEP} {\bf 08} (2019)
  143}, \href{http://arxiv.org/abs/1903.11123}{{\tt arXiv:1903.11123
  [hep-th]}}.

\bibitem{Jeong:2019pzg}
S.~Jeong, ``{SCFT/VOA correspondence via $\Omega$-deformation},''
  \href{http://dx.doi.org/10.1007/JHEP10(2019)171}{{\em JHEP} {\bf 10} (2019)
  171}, \href{http://arxiv.org/abs/1904.00927}{{\tt arXiv:1904.00927
  [hep-th]}}.

\bibitem{Gorsky:2017hro}
A.~Gorsky, B.~Le~Floch, A.~Milekhin, and N.~Sopenko, ``{Surface defects and
  instanton\textendash{}vortex interaction},''
  \href{http://dx.doi.org/10.1016/j.nuclphysb.2017.04.010}{{\em Nucl. Phys. B}
  {\bf 920} (2017)  122--156}, \href{http://arxiv.org/abs/1702.03330}{{\tt
  arXiv:1702.03330 [hep-th]}}.

\bibitem{Drukker:2012sr}
N.~Drukker, T.~Okuda, and F.~Passerini, ``{Exact results for vortex loop
  operators in 3d supersymmetric theories},''
  \href{http://dx.doi.org/10.1007/JHEP07(2014)137}{{\em JHEP} {\bf 07} (2014)
  137}, \href{http://arxiv.org/abs/1211.3409}{{\tt arXiv:1211.3409 [hep-th]}}.

\bibitem{Giombi:2009ek}
S.~Giombi and V.~Pestun, ``{The 1/2 BPS 't Hooft loops in N=4 SYM as instantons
  in 2d Yang-Mills},''
  \href{http://dx.doi.org/10.1088/1751-8113/46/9/095402}{{\em J. Phys. A} {\bf
  46} (2013)  095402}, \href{http://arxiv.org/abs/0909.4272}{{\tt
  arXiv:0909.4272 [hep-th]}}.

\bibitem{Giombi:2009ds}
S.~Giombi and V.~Pestun, ``{Correlators of local operators and 1/8 BPS Wilson
  loops on S**2 from 2d YM and matrix models},''
  \href{http://dx.doi.org/10.1007/JHEP10(2010)033}{{\em JHEP} {\bf 10} (2010)
  033}, \href{http://arxiv.org/abs/0906.1572}{{\tt arXiv:0906.1572 [hep-th]}}.

\bibitem{Assel:2015oxa}
B.~Assel and J.~Gomis, ``{Mirror Symmetry And Loop Operators},''
  \href{http://dx.doi.org/10.1007/JHEP11(2015)055}{{\em JHEP} {\bf 11} (2015)
  055}, \href{http://arxiv.org/abs/1506.01718}{{\tt arXiv:1506.01718
  [hep-th]}}.

\bibitem{Lamy-Poirier:2014sea}
J.~Lamy-Poirier, ``{Localization of a supersymmetric gauge theory in the
  presence of a surface defect},'' \href{http://arxiv.org/abs/1412.0530}{{\tt
  arXiv:1412.0530 [hep-th]}}.

\bibitem{Gukov:2013zka}
S.~Gukov and A.~Kapustin, ``{Topological Quantum Field Theory, Nonlocal
  Operators, and Gapped Phases of Gauge Theories},''
  \href{http://arxiv.org/abs/1307.4793}{{\tt arXiv:1307.4793 [hep-th]}}.

\bibitem{Gukov:2006jk}
S.~Gukov and E.~Witten, ``{Gauge Theory, Ramification, And The Geometric
  Langlands Program},'' \href{http://arxiv.org/abs/hep-th/0612073}{{\tt
  arXiv:hep-th/0612073}}.

\bibitem{Gaiotto:2012xa}
D.~Gaiotto, L.~Rastelli, and S.~S. Razamat, ``{Bootstrapping the superconformal
  index with surface defects},''
  \href{http://dx.doi.org/10.1007/JHEP01(2013)022}{{\em JHEP} {\bf 01} (2013)
  022}, \href{http://arxiv.org/abs/1207.3577}{{\tt arXiv:1207.3577 [hep-th]}}.

\bibitem{Gaiotto:2014ina}
D.~Gaiotto and H.-C. Kim, ``{Surface defects and instanton partition
  functions},'' \href{http://dx.doi.org/10.1007/JHEP10(2016)012}{{\em JHEP}
  {\bf 10} (2016)  012}, \href{http://arxiv.org/abs/1412.2781}{{\tt
  arXiv:1412.2781 [hep-th]}}.

\bibitem{Pan:2016fbl}
Y.~Pan and W.~Peelaers, ``{Intersecting Surface Defects and Instanton Partition
  Functions},'' \href{http://dx.doi.org/10.1007/JHEP07(2017)073}{{\em JHEP}
  {\bf 07} (2017)  073}, \href{http://arxiv.org/abs/1612.04839}{{\tt
  arXiv:1612.04839 [hep-th]}}.

\bibitem{Gomis:2016ljm}
J.~Gomis, B.~Le~Floch, Y.~Pan, and W.~Peelaers, ``{Intersecting Surface Defects
  and Two-Dimensional CFT},''
  \href{http://dx.doi.org/10.1103/PhysRevD.96.045003}{{\em Phys. Rev. D} {\bf
  96} (2017) no.~4, 045003}, \href{http://arxiv.org/abs/1610.03501}{{\tt
  arXiv:1610.03501 [hep-th]}}.

\bibitem{Nieri:2018ghd}
F.~Nieri, Y.~Pan, and M.~Zabzine, ``{Bootstrapping the $S^5$ partition
  function},'' \href{http://dx.doi.org/10.1051/epjconf/201819106005}{{\em EPJ
  Web Conf.} {\bf 191} (2018)  06005},
  \href{http://arxiv.org/abs/1807.11900}{{\tt arXiv:1807.11900 [hep-th]}}.

\bibitem{Nieri:2018pev}
F.~Nieri, Y.~Pan, and M.~Zabzine, ``{3d Mirror Symmetry from S-duality},''
  \href{http://dx.doi.org/10.1103/PhysRevD.98.126002}{{\em Phys. Rev. D} {\bf
  98} (2018) no.~12, 126002}, \href{http://arxiv.org/abs/1809.00736}{{\tt
  arXiv:1809.00736 [hep-th]}}.

\bibitem{Jeong:2020uxz}
S.~Jeong and N.~Nekrasov, ``{Riemann-Hilbert correspondence and blown up
  surface defects},'' \href{http://dx.doi.org/10.1007/JHEP12(2020)006}{{\em
  JHEP} {\bf 12} (2020)  006}, \href{http://arxiv.org/abs/2007.03660}{{\tt
  arXiv:2007.03660 [hep-th]}}.

\bibitem{Ferrari:2020avq}
F.~Ferrari and P.~Putrov, ``{Supergroups, q-series and 3-manifolds},''
  \href{http://arxiv.org/abs/2009.14196}{{\tt arXiv:2009.14196 [hep-th]}}.

\bibitem{Alday:2009aq}
L.~F. Alday, D.~Gaiotto, and Y.~Tachikawa, ``{Liouville Correlation Functions
  from Four-dimensional Gauge Theories},''
  \href{http://dx.doi.org/10.1007/s11005-010-0369-5}{{\em Lett. Math. Phys.}
  {\bf 91} (2010)  167--197}, \href{http://arxiv.org/abs/0906.3219}{{\tt
  arXiv:0906.3219 [hep-th]}}.

\bibitem{Gomis:2014eya}
J.~Gomis and B.~Le~Floch, ``{M2-brane surface operators and gauge theory
  dualities in Toda},'' \href{http://dx.doi.org/10.1007/JHEP04(2016)183}{{\em
  JHEP} {\bf 04} (2016)  183}, \href{http://arxiv.org/abs/1407.1852}{{\tt
  arXiv:1407.1852 [hep-th]}}.

\bibitem{Bonelli:2011wx}
G.~Bonelli, A.~Tanzini, and J.~Zhao, ``{The Liouville side of the Vortex},''
  \href{http://dx.doi.org/10.1007/JHEP09(2011)096}{{\em JHEP} {\bf 09} (2011)
  096}, \href{http://arxiv.org/abs/1107.2787}{{\tt arXiv:1107.2787 [hep-th]}}.

\bibitem{Alday:2009fs}
L.~F. Alday, D.~Gaiotto, S.~Gukov, Y.~Tachikawa, and H.~Verlinde, ``{Loop and
  surface operators in N=2 gauge theory and Liouville modular geometry},''
  \href{http://dx.doi.org/10.1007/JHEP01(2010)113}{{\em JHEP} {\bf 01} (2010)
  113}, \href{http://arxiv.org/abs/0909.0945}{{\tt arXiv:0909.0945 [hep-th]}}.

\bibitem{Nieri:2013vba}
F.~Nieri, S.~Pasquetti, F.~Passerini, and A.~Torrielli, ``{5D partition
  functions, q-Virasoro systems and integrable spin-chains},''
  \href{http://dx.doi.org/10.1007/JHEP12(2014)040}{{\em JHEP} {\bf 12} (2014)
  040}, \href{http://arxiv.org/abs/1312.1294}{{\tt arXiv:1312.1294 [hep-th]}}.

\bibitem{Aprile:2018oau}
F.~Aprile, S.~Pasquetti, and Y.~Zenkevich, ``{Flipping the head of $T[SU(N)]$:
  mirror symmetry, spectral duality and monopoles},''
  \href{http://dx.doi.org/10.1007/JHEP04(2019)138}{{\em JHEP} {\bf 04} (2019)
  138}, \href{http://arxiv.org/abs/1812.08142}{{\tt arXiv:1812.08142
  [hep-th]}}.

\bibitem{Benvenuti:2016wet}
S.~Benvenuti and S.~Pasquetti, ``{3d $ \mathcal{N} $ = 2 mirror symmetry,
  pq-webs and monopole superpotentials},''
  \href{http://dx.doi.org/10.1007/JHEP08(2016)136}{{\em JHEP} {\bf 08} (2016)
  136}, \href{http://arxiv.org/abs/1605.02675}{{\tt arXiv:1605.02675
  [hep-th]}}.

\bibitem{Cheng:2020zbh}
S.~Cheng, ``{Mirror Symmetry and Mixed Chern-Simons Levels for Abelian 3d N =
  2},'' \href{http://arxiv.org/abs/2010.15074}{{\tt arXiv:2010.15074
  [hep-th]}}.

\bibitem{Intriligator:1996ex}
K.~A. Intriligator and N.~Seiberg, ``{Mirror symmetry in three-dimensional
  gauge theories},'' \href{http://dx.doi.org/10.1016/0370-2693(96)01088-X}{{\em
  Phys. Lett. B} {\bf 387} (1996)  513--519},
  \href{http://arxiv.org/abs/hep-th/9607207}{{\tt arXiv:hep-th/9607207}}.

\bibitem{Kapustin:1999ha}
A.~Kapustin and M.~J. Strassler, ``{On mirror symmetry in three-dimensional
  Abelian gauge theories},''
  \href{http://dx.doi.org/10.1088/1126-6708/1999/04/021}{{\em JHEP} {\bf 04}
  (1999)  021}, \href{http://arxiv.org/abs/hep-th/9902033}{{\tt
  arXiv:hep-th/9902033}}.

\bibitem{Dimofte:2011py}
T.~Dimofte, D.~Gaiotto, and S.~Gukov, ``{3-Manifolds and 3d Indices},''
  \href{http://dx.doi.org/10.4310/ATMP.2013.v17.n5.a3}{{\em Adv. Theor. Math.
  Phys.} {\bf 17} (2013) no.~5, 975--1076},
  \href{http://arxiv.org/abs/1112.5179}{{\tt arXiv:1112.5179 [hep-th]}}.

\bibitem{Dimofte:2010tz}
T.~Dimofte, S.~Gukov, and L.~Hollands, ``{Vortex Counting and Lagrangian
  3-manifolds},'' \href{http://dx.doi.org/10.1007/s11005-011-0531-8}{{\em Lett.
  Math. Phys.} {\bf 98} (2011)  225--287},
  \href{http://arxiv.org/abs/1006.0977}{{\tt arXiv:1006.0977 [hep-th]}}.

\bibitem{Dimofte:2011ju}
T.~Dimofte, D.~Gaiotto, and S.~Gukov, ``{Gauge Theories Labelled by
  Three-Manifolds},'' \href{http://dx.doi.org/10.1007/s00220-013-1863-2}{{\em
  Commun. Math. Phys.} {\bf 325} (2014)  367--419},
  \href{http://arxiv.org/abs/1108.4389}{{\tt arXiv:1108.4389 [hep-th]}}.

\bibitem{Gukov:2016gkn}
S.~Gukov, P.~Putrov, and C.~Vafa, ``{Fivebranes and 3-manifold homology},''
  \href{http://dx.doi.org/10.1007/JHEP07(2017)071}{{\em JHEP} {\bf 07} (2017)
  071}, \href{http://arxiv.org/abs/1602.05302}{{\tt arXiv:1602.05302
  [hep-th]}}.

\bibitem{Nedelin:2016gwu}
A.~Nedelin, F.~Nieri, and M.~Zabzine, ``{$q$-Virasoro modular double and 3d
  partition functions},''
  \href{http://dx.doi.org/10.1007/s00220-017-2882-1}{{\em Commun. Math. Phys.}
  {\bf 353} (2017) no.~3, 1059--1102},
  \href{http://arxiv.org/abs/1605.07029}{{\tt arXiv:1605.07029 [hep-th]}}.

\bibitem{Nedelin:2015mio}
A.~Nedelin and M.~Zabzine, ``{q-Virasoro constraints in matrix models},''
  \href{http://dx.doi.org/10.1007/JHEP03(2017)098}{{\em JHEP} {\bf 03} (2017)
  098}, \href{http://arxiv.org/abs/1511.03471}{{\tt arXiv:1511.03471
  [hep-th]}}.

\bibitem{Lodin:2017lrc}
R.~Lodin, F.~Nieri, and M.~Zabzine, ``{Elliptic modular double and 4d partition
  functions},'' \href{http://dx.doi.org/10.1088/1751-8121/aa9a2d}{{\em J. Phys.
  A} {\bf 51} (2018) no.~4, 045402},
  \href{http://arxiv.org/abs/1703.04614}{{\tt arXiv:1703.04614 [hep-th]}}.

\bibitem{Nieri:2017vrb}
F.~Nieri, Y.~Pan, and M.~Zabzine, ``{$q$-Virasoro modular triple},''
  \href{http://dx.doi.org/10.1007/s00220-019-03371-1}{{\em Commun. Math. Phys.}
  {\bf 366} (2019) no.~1, 397--422},
  \href{http://arxiv.org/abs/1710.07170}{{\tt arXiv:1710.07170 [hep-th]}}.

\bibitem{Kim:2012gu}
H.-C. Kim, S.-S. Kim, and K.~Lee, ``{5-dim Superconformal Index with Enhanced
  En Global Symmetry},'' \href{http://dx.doi.org/10.1007/JHEP10(2012)142}{{\em
  JHEP} {\bf 10} (2012)  142}, \href{http://arxiv.org/abs/1206.6781}{{\tt
  arXiv:1206.6781 [hep-th]}}.

\bibitem{Kim:2013nva}
H.-C. Kim, S.~Kim, S.-S. Kim, and K.~Lee, ``{The general M5-brane
  superconformal index},'' \href{http://arxiv.org/abs/1307.7660}{{\tt
  arXiv:1307.7660 [hep-th]}}.

\bibitem{1993alg.geom..7001J}
L.~C. {Jeffrey} and F.~C. {Kirwan}, ``{Localization for nonabelian group
  actions},''{\em arXiv e-prints} (July, 1993)  alg--geom/9307001,
  \href{http://arxiv.org/abs/alg-geom/9307001}{{\tt arXiv:alg-geom/9307001
  [math.AG]}}.

\bibitem{Krattenthaler:2011da}
C.~Krattenthaler, V.~Spiridonov, and G.~Vartanov, ``{Superconformal indices of
  three-dimensional theories related by mirror symmetry},''
  \href{http://dx.doi.org/10.1007/JHEP06(2011)008}{{\em JHEP} {\bf 06} (2011)
  008}, \href{http://arxiv.org/abs/1103.4075}{{\tt arXiv:1103.4075 [hep-th]}}.

\bibitem{Awata:2008ed}
H.~Awata and H.~Kanno, ``{Refined BPS state counting from Nekrasov's formula
  and Macdonald functions},''
  \href{http://dx.doi.org/10.1142/S0217751X09043006}{{\em Int. J. Mod. Phys. A}
  {\bf 24} (2009)  2253--2306}, \href{http://arxiv.org/abs/0805.0191}{{\tt
  arXiv:0805.0191 [hep-th]}}.

\bibitem{Nekrasov:2003rj}
N.~Nekrasov and A.~Okounkov, ``{Seiberg-Witten theory and random partitions},''
  \href{http://dx.doi.org/10.1007/0-8176-4467-9_15}{{\em Prog. Math.} {\bf 244}
  (2006)  525--596}, \href{http://arxiv.org/abs/hep-th/0306238}{{\tt
  arXiv:hep-th/0306238}}.

\bibitem{Nekrasov:2002qd}
N.~A. Nekrasov, ``{Seiberg-Witten prepotential from instanton counting},''
  \href{http://dx.doi.org/10.4310/ATMP.2003.v7.n5.a4}{{\em Adv. Theor. Math.
  Phys.} {\bf 7} (2003) no.~5, 831--864},
  \href{http://arxiv.org/abs/hep-th/0206161}{{\tt arXiv:hep-th/0206161}}.

\bibitem{Tachikawa:2004ur}
Y.~Tachikawa, ``{Five-dimensional Chern-Simons terms and Nekrasov's instanton
  counting},'' \href{http://dx.doi.org/10.1088/1126-6708/2004/02/050}{{\em
  JHEP} {\bf 02} (2004)  050}, \href{http://arxiv.org/abs/hep-th/0401184}{{\tt
  arXiv:hep-th/0401184}}.

\bibitem{Sulkowski:2009ne}
P.~Sulkowski, ``{Matrix models for beta-ensembles from Nekrasov partition
  functions},'' \href{http://dx.doi.org/10.1007/JHEP04(2010)063}{{\em JHEP}
  {\bf 04} (2010)  063}, \href{http://arxiv.org/abs/0912.5476}{{\tt
  arXiv:0912.5476 [hep-th]}}.

\end{thebibliography}\endgroup
}

\end{document}